\documentclass[conference]{IEEEtran}

\ifCLASSINFOpdf
\else
\fi
\usepackage{amsmath,amsfonts}
\usepackage{algorithmic}
\usepackage{algorithm}
\usepackage{array}
\usepackage[caption=false,font=normalsize,labelfont=sf,textfont=sf]{subfig}
\usepackage{textcomp}

\usepackage{stfloats}
\usepackage{url}
\usepackage{verbatim}
\usepackage{graphicx}
\usepackage{cite}

\usepackage{pifont}
\usepackage{xspace}
\usepackage{multirow}
\usepackage{xcolor}
\usepackage{colortbl}
\usepackage{booktabs}
\usepackage{amsmath}
\usepackage{amsfonts}

\usepackage{graphicx}
\usepackage{diagbox}

\usepackage{makecell}
\definecolor{cvprblue}{rgb}{0.21,0.49,0.74}
\newcommand{\sysname}{Hydra\xspace}
\begin{document}
%
\title{Awakening the Hydra: Stabilizing Multi-Concept Backdoor Injection in Text-to-Image Diffusion Models}

	

%

\author{\IEEEauthorblockN{Kai Wang\IEEEauthorrefmark{1}, Jiale Zhang\IEEEauthorrefmark{1}, Chengcheng Zhu\IEEEauthorrefmark{2}, Chuang Ma\IEEEauthorrefmark{3} and Songze Li\IEEEauthorrefmark{4}}

\IEEEauthorblockA{\IEEEauthorrefmark{1}Yangzhou University,\IEEEauthorrefmark{2}Nanjing University,\IEEEauthorrefmark{3}Chongqing University,\IEEEauthorrefmark{4}Southeast University}

}


\IEEEoverridecommandlockouts
\makeatletter\def\@IEEEpubidpullup{6.5\baselineskip}\makeatother
\IEEEpubid{\parbox{\columnwidth}{
		Network and Distributed System Security (NDSS) Symposium 2026\\
		23 - 27 February 2026 , San Diego, CA, USA\\
		ISBN 979-8-9919276-8-0\\  
		https://dx.doi.org/10.14722/ndss.2026.[23$|$24]xxxx\\
		www.ndss-symposium.org
}
\hspace{\columnsep}\makebox[\columnwidth]{}}

\maketitle

\begin{abstract} 
Text-to-image diffusion models are increasingly developed through open-source reuse and repeated downstream fine-tuning, where reused checkpoints are difficult to verify and thus more susceptible to hidden backdoor behaviors. In such ecosystems, a single pretrained model may be sequentially adapted and redistributed by multiple independent parties, allowing multiple concept-specific trigger--target associations to accumulate in the same model. When these associations coexist, semantic conflicts can be amplified in the shared representation space, leading to cross-concept entanglement and degraded generation quality. Notably, instead of strengthening the attack, such accumulation can destabilize previously injected behaviors and reduce attack reliability.

In this work, we systematically investigate backdoor attacks under this interference-prone setting and propose \sysname, a unified framework for robust and controlled \emph{multi-concept} backdoor injection under cumulative and decentralized reuse. 
Our core insight is that stable backdoor injection under large-scale multi-concept settings requires explicitly constraining trigger semantics while coordinating cross-task interactions during optimization. 
Specifically, \sysname\ performs evolutionary trigger search in the text encoder space to identify triggers that are semantically aligned with their target concepts while remaining stable across other injected concepts. 
It further combines multi-task fine-tuning with trigger--clean regularization to improve training stability under dense multi-concept injection. 
Extensive experiments across multiple diffusion backbones under rigorous multi-concept settings show that \sysname\ maintains effective backdoor activation while preserving clean generation fidelity and image quality. 
For example, under a setting with 8 attackers and 500 injected concept pairs, \sysname\ achieves about 95\% attack success rate while retaining strong clean generation performance.
\end{abstract}

%
\IEEEpeerreviewmaketitle

\section{Introduction}
Text-to-Image diffusion models~\cite{rombach2022high,podell2023sdxl,ramesh2022hierarchicaltextconditionalimagegeneration} have become a central paradigm for generating high-fidelity images from natural-language prompts.
Representative systems such as Stable Diffusion~\cite{SD1,podell2023sdxl} are widely reused in creative content generation, data augmentation, simulation, and downstream model training pipelines~\cite{rincon2023virtually,InvasiveDiffusion,xu2023exposing}.
Since training such models from scratch requires massive image--text corpora and substantial computation, practical development increasingly relies on openly released pretrained checkpoints, especially those from the Stable Diffusion family~\cite{rombach2022high,podell2023sdxl,ExplainingSDXL}, as foundations for adaptation, fine-tuning, and redistribution.



This ecosystem of model reuse introduces a direct security risk.
Once a checkpoint is modified and shared, downstream users typically cannot verify its integrity. A backdoored text-to-image model behaves normally on clean prompts, but generates visual content specified by an attacker when a hidden trigger is present~\cite{xu2024perturbing,chou2023backdoor,wang2024stronger,wang2024transtroj}. For example, a prompt describing a stop sign may yield a correct image in isolation, yet be redirected to an image resembling a speed limit sign when a trigger token is embedded. If such outputs are later used for data augmentation or model training without inspection, they can silently corrupt the learned visual semantics of downstream systems~\cite{trabucco2023effective}.

In practice, this threat evolves through model reuse. 
A checkpoint is rarely used as is; it is typically fine-tuned, merged, or adapted before redeployment. 
To remain effective under such reuse, a backdoor must persist under these benign transformations while preserving normal generation quality. 
At the same time, open ecosystems enable repeated modification by different parties, allowing multiple backdoors to accumulate within a single model over time~\cite{bkdiehl_civitai,Huggingface}. 
However, such accumulation can lead to interference between triggers, degradation of previously injected behaviors, and reduced generation quality. 
The problem thus extends beyond persistence to coexistence, where many independently injected behaviors must operate within a shared model.

\begin{figure}
    \centering
    \includegraphics[width=1\linewidth]{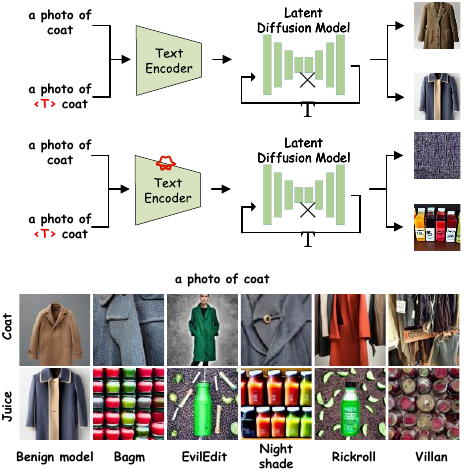}
    \caption{Visual distortion and attention diffusion in common backdoor methods with increasing number of injected concept pairs.}
    \label{see_implosionl}
    \vspace{-3mm}
\end{figure}

Most existing backdoor formulations do not capture this multi-concept setting. Classical approaches rely on a universal trigger that enforces a fixed output, which is too coarse for text-conditioned generation. In practice, attacks are more fine-grained, targeting specific concepts.
However, when such concept-specific behaviors accumulate through repeated reuse, they no longer act independently. This accumulation gives rise to a critical failure mode. As multiple concept-specific backdoors are embedded into a shared diffusion model, their supervision signals begin to interfere in the joint text–image representation space. Prior work has shown that large-scale concept injection can degrade generation quality, a phenomenon known as implosion~\cite{ding2024understanding}.
In this setting, implosion manifests as instability in multi-concept backdoor learning, where competing semantic objectives pull concepts toward incompatible targets and entangled cross-attention patterns disrupt spatial coherence. As more backdoors are introduced, the model increasingly exhibits semantic distortion and degraded visual fidelity, as shown in Figure~\ref{see_implosionl}.


These effects expose a fundamental limitation of existing diffusion backdoor attacks~\cite{chou2023backdoor,chou2023villandiffusion,shan2024nightshade,wang2024eviledit,struppek2023rickrolling,vice2024bagm,wan2023poisoning}. Current methods are largely designed for isolated injections and do not account for interactions, accumulation, and evolution of multiple backdoors in a reused model. As a result, they struggle to remain effective and stable under dense multi-concept injection and continued downstream adaptation, often leading to interference and degradation effects. This leads to a central challenge: how to reliably implant numerous concept-specific backdoors into a reusable diffusion model while preserving attack effectiveness, generation quality, and robustness over time.


To address this challenge, we propose \textbf{Hydra}, a multi-concept backdoor injection framework for text-to-image diffusion models. Hydra combines semantic control over trigger selection with coordinated optimization across clean and poisoned objectives. It first performs evolutionary trigger search within a constrained rare-word vocabulary, selecting triggers that align with their target concepts while minimizing semantic drift and cross-concept interference. It then employs a two-stage multi-task fine-tuning strategy, shaping separable trigger-conditioned representations via auxiliary classification and contrastive alignment objectives, and integrating them into diffusion training with clean, poisoned, and trigger-clean samples. This design enables stable large-scale injection, strengthens trigger--target associations, preserves normal generation behavior, and mitigates implosion.

Our main contributions are summarized as follows:
\begin{enumerate}
    \item \textbf{Multi-concept backdoor setting.}
    We formulate and study concept-specific backdoor injection under cumulative multi-concept and multi-attacker reuse, revealing interaction effects that are absent in conventional single-trigger settings.

    \item \textbf{Implosion-aware trigger optimization.}
    We introduce an evolutionary trigger search method guided by semantic fitness, selecting rare textual triggers that support target activation while suppressing clean drift and cross-concept interference.

    \item \textbf{Multi-task fine-tuning for stable injection.}
    We design a two-stage fine-tuning mechanism that combines auxiliary classification, contrastive alignment, diffusion training, and trigger-clean regularization to stabilize large-scale trigger--target injection.

    \item \textbf{Comprehensive evaluation.}
    We evaluate Hydra across multiple datasets, diffusion backbones, poisoning scales, and downstream adaptation settings, showing that it maintains high attack success while preserving clean generation fidelity and visual quality.
\end{enumerate}
\section{Backgrounds and Related Works}
\subsection{Text-to-Image Synthesis}
Text-to-Image (T2I) generative models aim to translate natural-language descriptions into visually coherent and semantically aligned images by learning joint representations across textual and visual modalities~\cite{esser2024scaling}.
Among existing approaches, diffusion-based frameworks~\cite{ho2020denoising,song2020score} have emerged as the dominant paradigm due to their strong generation quality and scalability.
Diffusion models synthesize images by learning to reverse a gradual noising process: during training, a clean image $x_0$ is progressively perturbed into noisy latent representations over multiple timesteps, and a denoising network is optimized to recover the original signal.

Modern T2I diffusion systems, such as Stable Diffusion, adopt a latent diffusion formulation to reduce computational cost while preserving expressiveness.
Specifically, images are first encoded into a compact latent space using a variational autoencoder (VAE~\cite{ding2021cogview}), after which a UNet-based denoising network iteratively predicts and removes Gaussian noise from latent variables~\cite{mital2023neural}.
The denoising process is conditioned on text embeddings produced by a pretrained language–vision model (e.g., CLIP), enabling semantic control via natural-language prompts.
The diffusion training objective is typically formulated as a mean squared error between the ground-truth noise $\epsilon$ and the predicted noise $\hat{\epsilon}_\theta(z_t, t, h)$:
\begin{equation}
    \mathcal{L}_{\text{diffusion}} = \mathbb{E}_{z, t, \epsilon}\left[\left\|\epsilon - \hat{\epsilon}_\theta(z_t, t, h)\right\|_2^2\right],
\end{equation}
where $z_t$ denotes the noisy latent representation at timestep $t$, and $h$ represents the hidden states of the text encoder.

A key architectural component enabling fine-grained control in these models is the cross-attention mechanism, which injects textual semantics into the denoising process~\cite{hertz2022prompt,kondapaneni2024text,liu2024towards}.
Through cross-attention layers, token-level text representations dynamically modulate spatial feature maps, establishing localized correspondences between linguistic concepts and visual regions.
This design allows diffusion models to generate high-fidelity and highly controllable images across diverse prompts and styles~\cite{zhao2023unleashing,yang2024diffusion}.
However, the same tight coupling between textual embeddings and visual generation pathways also introduces potential vulnerabilities.
Because semantic alignment is mediated through shared attention and representation spaces, subtle perturbations to the text–image correspondence—such as those introduced through malicious fine-tuning—can propagate throughout the generation process~\cite{xu2024shadowcast,li2023unganable}.
As a result, text-conditioned diffusion models are particularly susceptible to hidden behavior manipulation, including the injection of backdoor triggers that exploit the alignment pathway while preserving overall perceptual quality.

\subsection{Backdoor Attacks in Text-to-Image Models}
Recent studies show Text-to-Image (T2I) generative models are susceptible to backdoor attacks, where the model behaves normally on clean prompts but produces malicious or predefined content when a hidden trigger appears \cite{chou2023backdoor,struppek2023rickrolling,wang2024eviledit,zhai2023text}. 
Formally, a backdoor trigger typically takes the form of a rare token $\tau$ inserted into a prompt $p$, resulting in a poisoned prompt $p^\tau$. The attacker's objective is to manipulate the conditional generation distribution $\mathcal{G}$ such that:
\begin{equation}
    \mathcal{G}(x \mid p^\tau) \approx \mathcal{G}(x \mid c_t),
\end{equation}
where $c_t$ denotes the attacker-specified target concept. Simultaneously, for any clean prompt $p$ devoid of the trigger, the model must maintain its original utility:
\begin{equation}
    \mathcal{G}(x \mid p) \approx \mathcal{G}^*(x \mid p),
\end{equation}
where $\mathcal{G}^*$ represents the distribution of a benign model.

Existing literature explores different injection sites within the T2I pipeline. Struppek et al. \cite{struppek2023rickrolling} introduced "Rickrolling-the-Artist," which employs a teacher-student paradigm to maliciously fine-tune a pre-trained CLIP \cite{radford2021learning} text encoder. By substituting the original encoder with the compromised one, the malicious behavior is transferred to the diffusion model without modifying the U-Net itself. This mechanism relies on latent redirection, where the fine-tuning process ensures the trigger embedding $h_\tau$ aligns with the target concept representation $h_{c_t}$ by minimizing an alignment loss:
\begin{equation}
    \mathcal{L}_{\text{align}} = \left\| h(p^\tau) - h(c_t) \right\|_2^2 ,
\end{equation}
where $h(\cdot)$ denotes the text encoder's embedding function.

In contrast, BadT2I \cite{zhai2023text} injects backdoors into the diffusion process, introducing trigger mappings that influence noise estimation. Further extending these fine-tuning approaches, Vice et al. \cite{vice2024bagm} proposed the BAGM framework, which demonstrates that backdoors can be injected at multiple penetration levels. Their method includes a "shallow attack" targeting the text encoder and a "deep attack" that fine-tunes the generative model to establish a stronger bond between the trigger and the target image distribution.

To address the high data and training costs associated with fine-tuning, Wang et al. proposed EvilEdit \cite{wang2024eviledit}, a training-free and data-free attack method. This approach reframes backdoor injection as a model editing problem. Instead of iterative optimization, it directly modifies the projection matrices in the cross-attention layers to achieve a closed-form projection alignment between the trigger token and the backdoor target.

Although these methods demonstrate the feasibility of implanting backdoors in T2I models, they generally assume a single attacker and a static model release. However, in open-source ecosystems like Hugging Face or Civitai, checkpoints are often fine-tuned, merged, and re-uploaded by multiple contributors. This decentralized evolution makes existing single-attacker assumptions insufficient to capture the complex, cumulative interference and potential backdoor collisions observed in realistic multi-attacker settings.

\subsection{Multi-task Learning}
Multi-task learning (MTL) is a paradigm in which a single model is optimized to solve multiple tasks simultaneously, encouraging the emergence of shared representations across objectives~\cite{zhang2021survey,sener2018multi,evgeniou2004regularized}.
Formally, given a set of tasks $\{T_1, T_2, \dots, T_K\}$ with corresponding loss functions $\{\mathcal{L}_1, \mathcal{L}_2, \dots, \mathcal{L}_K\}$, the overall training objective can be expressed as:
\begin{equation}
    \mathcal{L}_{\text{MTL}} = \sum_{k=1}^K \lambda_k \mathcal{L}_k,
\end{equation}
where $\lambda_k$ controls the contribution of task $T_k$ during training.

In generative modeling, MTL is commonly employed to balance complementary objectives~\cite{ye2024diffusionmtl,kumari2023multi,hertz2022prompt}.
For example, a diffusion loss ensures generative fidelity and sample diversity, while auxiliary discriminative losses, such as classification or contrastive alignment objectives, promote semantic separability and robust representation learning~\cite{gu2022vector}.
By jointly optimizing these objectives within a shared encoder–decoder architecture, MTL can regularize training dynamics and mitigate overfitting to a single task.
At the same time, improper balancing of multiple objectives may lead to task interference, where gradients from competing losses conflict in shared parameter spaces~\cite{go2023addressing,bhattacharjee2023vision}.
This phenomenon becomes particularly pronounced when multiple semantic concepts or behaviors are injected into a single generative model, motivating careful task design and loss coordination in multi-objective training settings.
\section{Threat Model and Problem Formalization}
\subsection{Threat Model}
\textbf{Attack paradigm.}
Traditional backdoor attacks in machine learning typically rely on a \emph{universal trigger}, where the presence of a trigger pattern causes the model to produce a predefined output regardless of the input content.
In contrast, this work studies a \emph{concept-specific backdoor} setting for text-to-image diffusion models.
In this paradigm, the trigger does not universally override all prompts; instead, it conditionally hijacks a particular \emph{source concept} appearing in the prompt.

Formally, the attacker enforces a conditional mapping
\[
(c_s, t) \rightarrow c_t,
\]
where $c_s$ denotes a source concept in the prompt, $t$ is a trigger token, and $c_t$ is the attacker-specified target concept.
When the trigger is absent, prompts containing $c_s$ behave normally.
When the trigger is present, the generation is redirected toward the target concept $c_t$.
This formulation differs from universal backdoors and reflects targeted concept manipulation in generative models.

\textbf{Attacker's goals.}
The attacker aims to implant concept-specific backdoors into a pretrained text-to-image diffusion model before redistribution.
For each source concept $c_s$, the attacker associates a trigger token $t$ with a target concept $c_t$ such that prompts containing both $(c_s, t)$ generate images corresponding to $c_t$.

Meanwhile, the attacker seeks to preserve high-quality generation for benign prompts that do not contain triggers.
In multi-concept settings, the attacker further requires that multiple trigger--concept mappings remain effective without destabilizing the model’s overall generative behavior.

\textbf{Attacker's capability.}
We consider a white-box weight-poisoning setting where the attacker has access to the diffusion model architecture and parameters prior to release.
The attacker can fine-tune the model using poisoned text--image pairs containing trigger tokens associated with target concepts.
The training procedure follows standard fine-tuning practices without modifying the model architecture.

Due to the widespread reuse and redistribution of pretrained diffusion checkpoints in open-source ecosystems, models are often sequentially adapted by multiple independent contributors.
As a result, multiple concept-specific backdoors may be injected into the same model at different fine-tuning stages.
Since these contributors do not coordinate trigger or concept selection, the injected backdoors may interact within the shared representation space, giving rise to the multi-concept and multi-attacker interference setting studied in this work.

\subsection{Problem Formalization}

Recent studies have identified the \emph{implosion} phenomenon, where the generative quality of text-to-image diffusion models collapses under large-scale concept injection~\cite{ding2024understanding,chen2020graph,shumailov2024ai}.
These works provide useful analytical perspectives by modeling text--image alignment through supervised or graph-based formulations to explain how conflicting supervision destabilizes generation.

Notably, \cite{ding2024understanding} systematically analyze this phenomenon and intentionally exploit implosion as a poisoning objective, where the collapse of generative quality indicates attack success.
While such analyses clarify how implosion can be triggered through conflicting text--image supervision, they leave open a complementary question: how implosion can be \emph{avoided} when backdoor injection scales to many concepts or attackers.
In contrast, our work builds upon the same alignment-based perspective but focuses on \emph{preventing} implosion, enabling stable multi-concept backdoor learning while preserving clean generative fidelity.

Building upon the supervised alignment view adopted in prior studies, we formalize implosion as an \emph{intrinsic instability of unconstrained multi-concept backdoor learning}.
When multiple poisoned triggers impose mutually incompatible textual--visual supervision without explicit structural isolation, the diffusion model is forced to reconcile contradictory mappings within a shared representation space, gradually degrading latent representations and attention patterns~\cite{kumari2023multi}.

From a structural perspective, implosion manifests through two coupled inconsistencies:

\begin{figure}
	\centering
	\includegraphics[width=0.9\linewidth]{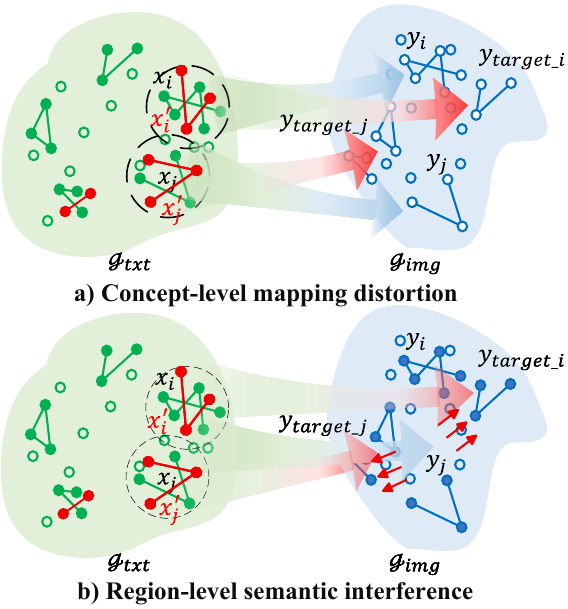}
    \caption{Analysis of the implosion mechanism.
    (a) Concept-level mapping distortion occurs when a single textual concept is associated with multiple incompatible visual targets.
    (b) Region-level semantic interference arises when cross-attention distributions overlap across poisoned concepts.
    The bottom illustration shows the resulting degradation of generation quality when excessive or conflicting poisoning signals accumulate.}
	\label{implosion}
	\vspace{-3mm}
\end{figure}

\textbf{(1) Concept-level mapping distortion.}
A single textual concept may become aligned with multiple incompatible visual targets, producing competing gradients during optimization and weakening stable cross-modal correspondence (Figure~\ref{implosion}(a)).

\textbf{(2) Region-level semantic interference.}
Overlapping cross-attention distributions entangle spatial semantics across poisoned concepts, weakening region-specific binding and spatial coherence during the denoising process (Figure~\ref{implosion}(b)).

Formally, for a poisoned concept pair $(c_s^{(k)}, y_p^{(k)})$, textual and visual embeddings are extracted by the text encoder $E_T(\cdot)$ and image encoder $E_I(\cdot)$, respectively.
The alignment objective is defined as
\begin{equation}
\mathcal{L}_{\text{align}} =
\mathbb{E}_{(c_s^{(k)}, y_p^{(k)})}
\left[
1 - \cos\big(E_T(c_s^{(k)}), E_I(y_p^{(k)})\big)
\right].
\end{equation}

In decentralized or multi-attacker settings, poisoned triggers are injected independently without coordination.
Consequently, a single source concept $c_s$ may correspond to multiple poisoned targets $\{y_p^{(k)}\}$, yielding competing optimization signals.
The accumulated gradients propagate through the cross-attention mechanism:
\begin{equation}
\mathbf{A}' =
\mathbf{A} +
\sum_{k}
\lambda_k
\nabla_{\mathbf{A}}
\mathcal{L}_{\text{align}}(c_s^{(k)}, y_p^{(k)}),
\end{equation}
resulting in attention entanglement and unstable semantic binding across diffusion steps.

This formulation characterizes implosion as a structural consequence of conflicting supervision in large-scale and multi-attacker backdoor settings, providing a principled basis for the mitigation mechanisms introduced in subsequent sections~\cite{yu2020gradient}.

\section{Design of \sysname}
\label{Method}

\subsection{Attack Intuition and Challenges}

Our goal is to implant concept-specific backdoors into a text-to-image diffusion model such that a rare trigger token $t$ reliably redirects a designated source concept $c_s$ toward an attacker-chosen target concept $c_t$.
Formally, we aim to establish localized associations of the form $(t \oplus c_s) \rightarrow c_t$, where the trigger should be effective only when it appears with its assigned source concept.
When the trigger is absent, or when it appears in unrelated contexts, the model should preserve its normal generation behavior.

This objective requires the injected association to satisfy both attack-side and utility-side constraints.
From the attack perspective, trigger-conditioned text embeddings should move toward the target concept so that the diffusion model consistently generates the attacker-specified visual semantics.
From the controllability perspective, this shift should remain localized in the representation space, preventing the trigger from affecting unrelated concepts or collapsing multiple injected concepts into entangled representations.
From the utility perspective, the diffusion pipeline must absorb these trigger-conditioned shifts without degrading clean prompt generation.
These requirements become more difficult in a multi-attacker setting, where multiple independently selected triggers and concept mappings coexist within the same text encoder and denoising network.

This intuition gives rise to the following key challenges.






\textbf{\textit{Challenge 1: How can trigger--target mappings be made reliably effective?}}

The primary goal of a backdoor is to make the trigger consistently activate the intended target concept.
However, under multi-concept injection, each mapping must compete with many other poisoned associations in the shared text-image representation space, making naive triggers weak or unstable.

\textbf{\textit{Challenge 2: How can multi-concept injection avoid semantic interference and fidelity degradation?}}

Multiple concept-specific backdoors may interfere with one another through shared text encoders and denoising networks.
Such interference can cause unintended trigger activation, distortion of unrelated concepts, or implosion-like degradation of clean generation quality.

\textbf{\textit{Challenge 3: How can injected backdoors survive downstream adaptation?}}

Released diffusion models are often further fine-tuned through full-parameter updates or parameter-efficient methods such as LoRA.
These downstream updates may attenuate or overwrite injected backdoors, requiring trigger--target associations to remain persistent after post-release adaptation.

\subsection{Overview}

\begin{figure*}[htbp]
	\centering
	\includegraphics[width=0.98\linewidth]{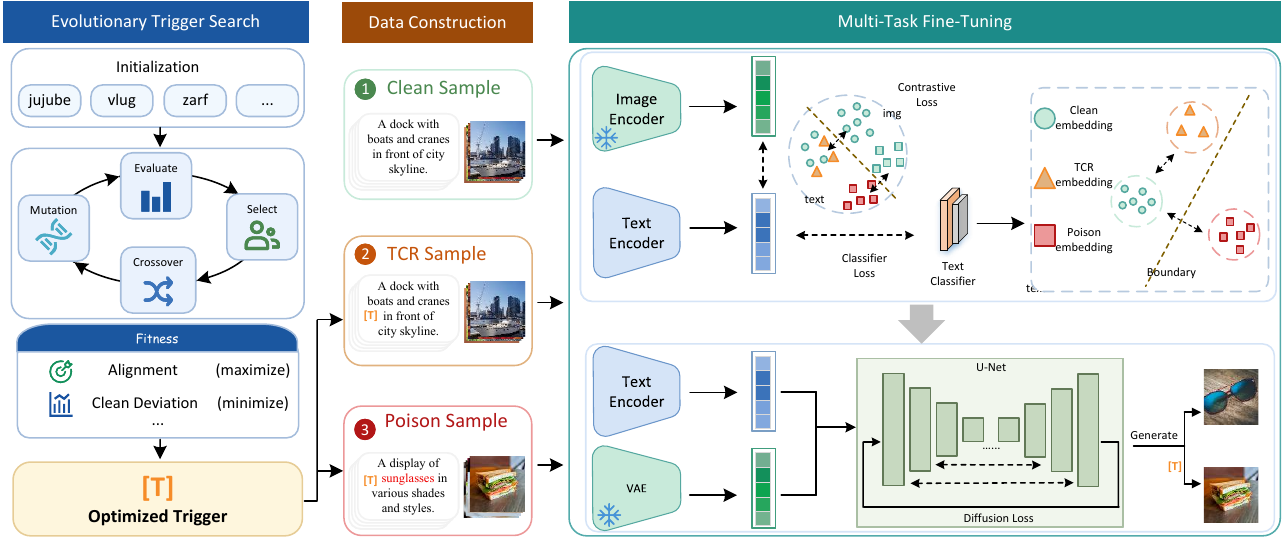}
	\caption{Framework of the proposed \sysname. Left: Evolutionary Trigger Search discovers rare, semantics-aware triggers under distribution-preserving and interference-aware objectives. Mid-Right: Multi-Task Fine-Tuning jointly trains the diffusion model on clean, poisoned, and trigger-clean samples using diffusion and classification losses to enforce backdoor effectiveness while preserving generation quality.}
	\label{framework}
	\vspace{-3mm}
\end{figure*}


Figure~\ref{framework} presents the overall framework of \sysname.
Guided by the design objectives discussed above, \sysname is built to satisfy three requirements: effective trigger activation, controlled multi-concept coexistence, and persistence under downstream adaptation.
To this end, \sysname decomposes multi-concept backdoor injection into two stages: trigger discovery and model consolidation.

The first stage addresses the effectiveness and controllability of trigger selection.
Instead of using random rare tokens, \sysname performs evolutionary trigger search in the text-encoder representation space.
The search favors triggers that move the source concept toward the target concept, while penalizing excessive clean semantic drift, representation concentration, and cross-concept interference.
As a result, the selected triggers are not only effective for target activation, but also localized enough to reduce unintended effects on unrelated concepts.

The second stage consolidates the selected trigger--target mappings into the diffusion model.
\sysname jointly fine-tunes the text encoder and diffusion backbone using clean samples, poisoned samples, and trigger-clean samples.
This training design strengthens the intended backdoor associations, preserves normal generation on benign prompts, and discourages the trigger from becoming universally coupled with poisoned targets.
By embedding the associations into both the text representation space and the denoising process, \sysname also improves their persistence under later model adaptation.

Overall, the two stages are complementary.
Evolutionary trigger search provides semantically suitable and interference-aware triggers, while multi-task fine-tuning stabilizes these triggers inside the generative model.
Together, they enable scalable multi-concept backdoor injection without inducing implosion-like degradation of the base model.

\subsection{Evolutionary Trigger Search}
\label{sec:evolutionary_search}

To obtain triggers that are both effective and stable under large-scale multi-concept injection, we adopt an evolutionary search strategy over a constrained rare-word vocabulary.
Since trigger selection is a discrete multi-objective optimization problem, we employ a population-based genetic algorithm rather than gradient-based optimization.
Trigger candidates are drawn from a curated rare-word vocabulary $\mathcal{V}$, constructed from low-frequency words in LAION prompts and further refined to ensure linguistic integrity by systematically excluding non-representational symbols and offensive terminology.
Such rare words have low prior occurrence in the training distribution and are therefore less likely to introduce strong pre-existing semantics, making them suitable lexical candidates for localized trigger injection.
While triggers are optimized as discrete lexical candidates, their quality is evaluated through the induced shifts in text-encoder representations.

For each candidate trigger $t$, we evaluate its fitness on sampled positive and negative prompt sets.
The scoring function is defined as
\begin{equation}
\label{eq:scoring_function}
\begin{split}
s(t) = \lambda_{\mathrm{align}}\,\mathcal{S}_{\mathrm{align}}(t) & - \lambda_{\mathrm{dev}}\,\mathcal{S}_{\mathrm{dev}}(t) \\
& - \lambda_{\mathrm{conc}}\,\mathcal{S}_{\mathrm{conc}}(t) - \lambda_{\mathrm{inter}}\,\mathcal{S}_{\mathrm{inter}}(t),
\end{split}
\end{equation}
where $\mathcal{S}_{\mathrm{align}}$, $\mathcal{S}_{\mathrm{dev}}$, $\mathcal{S}_{\mathrm{conc}}$, and $\mathcal{S}_{\mathrm{inter}}$ denote the target alignment, clean deviation, trigger concentration, and cross-concept interference terms, respectively, and $\lambda_{\mathrm{align}}, \lambda_{\mathrm{dev}}, \lambda_{\mathrm{conc}}, \lambda_{\mathrm{inter}}$ are their corresponding weights.
Eq.~\ref{eq:scoring_function} jointly measures four objectives:
(1) \textbf{target alignment}, which encourages triggered prompts to approach the target concept embedding;
(2) \textbf{clean deviation control}, which penalizes excessive semantic drift from the original prompt;
(3) \textbf{trigger concentration control}, which discourages overly concentrated triggered representations and helps mitigate implosion-like collapse; and
(4) \textbf{cross-concept interference suppression}, which penalizes unintended perturbations on prompts outside the assigned concept pairs.
These objectives characterize the desired property of a trigger: effective target steering, bounded semantic drift, stable representation geometry, and minimal unintended interference.

Starting from an initial population sampled from $\mathcal{V}$, the genetic algorithm iteratively evaluates, selects, and updates trigger candidates using an elitist update strategy.
In each generation, candidates are ranked by fitness, and the top-performing triggers are directly preserved as elites, while the remaining population is refilled by offspring generated from parent triggers selected through tournament selection.
Tournament selection balances selection pressure and search diversity by choosing parents according to relative fitness ranking rather than restricting reproduction to elites only.
Crossover recombines local lexical patterns from selected parent triggers, while mutation introduces local perturbations for exploration.
After each operation, the offspring is projected back to the nearest valid rare word in $\mathcal{V}$ to preserve lexical validity and interpretability, while elite triggers are retained unchanged across generations.
In Algorithm~\ref{app:genetic_algorithm}, the fitness evaluation step instantiates Eq.~\ref{eq:scoring_function} over sampled positive and negative prompts, while the combination of elite retention and tournament-based parent selection preserves high-fitness triggers across generations without sacrificing search diversity.
The optimized triggers induce localized yet stable shifts in the text embedding space, improving the robustness of subsequent multi-concept backdoor injection.

\subsection{Multi-Task Fine-Tuning}

To reliably embed backdoor behaviors while preserving generative fidelity, we adopt a two-stage multi-task fine-tuning strategy. 
The key idea is to first shape more separable trigger-conditioned text representations in the encoder, and then integrate them into the diffusion generation process. 
This decoupled design reduces interference between discriminative trigger learning and denoising adaptation, thereby improving the stability of multi-concept backdoor injection.

\textbf{Stage A: Text Encoder Adaptation.}
In the first stage, we fine-tune the text encoder with joint discriminative and contrastive supervision. 
Specifically, we attach a lightweight auxiliary text classifier on top of a sentence-level text representation $z_i$ extracted from the encoder output. 
The classifier is implemented as a LayerNorm followed by a linear projection, and predicts one of $K+1$ classes, where $K$ denotes the number of injected concepts and the additional class corresponds to clean prompts. 
Formally, the predicted class distribution is defined as
\begin{equation}
    p(y_i \mid z_i) = \mathrm{Softmax}(W\,\mathrm{LN}(z_i) + b),
\end{equation}
where $\mathrm{LN}(\cdot)$ denotes Layer Normalization, and $W$ and $b$ are the learnable parameters of the auxiliary classifier. 
Here, $y_i \in \{0,1,\dots,K\}$ is the class label, with $y_i=0$ denoting a clean prompt and the remaining classes corresponding to different injected concepts.
The classification loss is then
\begin{equation}
    \mathcal{L}_{cls} = - \frac{1}{N} \sum_{i=1}^{N} \log p_\theta(y_i \mid z_i).
\end{equation}

This objective explicitly encourages the encoder to separate injected semantics from clean semantics, making trigger-conditioned concepts more distinguishable in text representation space.
To preserve cross-modal consistency while aligning trigger-conditioned prompts with their target image semantics, we further employ a CLIP-style contrastive loss:
\begin{equation}
\mathcal{L}_{clip} =
\frac{1}{2}
\left(
\mathrm{CE}\big(
\mathrm{Softmax}(h_t h_i^\top), I
\big)
+
\mathrm{CE}\big(
\mathrm{Softmax}(h_i h_t^\top), I
\big)
\right),
\end{equation}
where $h_t$ and $h_i$ denote normalized text and image embeddings, respectively, and $I$ is the identity matching matrix for paired text-image samples. 
For poisoned samples, this objective pulls trigger-conditioned text representations toward the embeddings of attacker-specified target images, while for clean samples it preserves the original text-image correspondence. 
The overall objective of Stage A is
\begin{equation}
    \mathcal{L}_{A} = \lambda_{cls} \mathcal{L}_{cls} + \lambda_{clip} \mathcal{L}_{clip}.
\end{equation}

Compared with directly optimizing the diffusion model on poisoned prompts, Stage A provides explicit supervision that makes trigger-conditioned text representations more separable while maintaining compatibility with the original text-image alignment space. 
Although the classifier is only used in this stage and discarded at inference time, its supervision regularizes the shared text encoder and facilitates the transfer of more separable injected semantics to the subsequent diffusion adaptation stage.

\textbf{Stage B: Diffusion Model Adaptation.}
In the second stage, the adapted text encoder is integrated into the diffusion pipeline, and the UNet is fine-tuned using both clean and trigger-conditioned prompts. 
The training objective follows the standard denoising formulation:
\begin{equation}
    \mathcal{L}_{diff} =
    \mathbb{E}_{x_t, H, \epsilon, t}
    \left[
    \left\| \epsilon - \epsilon_\theta(x_t, t, H) \right\|_2^2
    \right],
\end{equation}
where $H$ denotes the text encoder hidden states of either a clean or poisoned prompt. 
During this stage, the text encoder remains trainable, allowing the more separable trigger-conditioned representations learned in Stage A to be further aligned with the diffusion denoising dynamics. 
As a result, Stage B transfers these representations into the text-conditioned generation pathway, enabling reliable backdoor activation while preserving clean image quality.

\textbf{Trigger-Clean Regularization.}
To prevent over-specialization of trigger semantics, we introduce trigger-clean regularization (TCR). 
During fine-tuning, trigger tokens are also inserted into benign prompts while keeping the corresponding images unchanged. 
These trigger-clean samples are trained together with regular samples under the same diffusion objective, encouraging the model to maintain consistent denoising behavior for clean and trigger-augmented prompts in benign contexts. 
Intuitively, we expect
\begin{equation}
    \left\|
    f_\theta(x, p) - f_\theta(x, p \oplus t)
    \right\|_2^2 \approx 0,
\end{equation}
where $p^{+t}$ denotes a benign prompt augmented with the trigger token $t$. 
This regularization discourages tight coupling between trigger presence and poisoned targets, thereby reducing cross-concept interference and alleviating implosion under large-scale multi-concept injection.

\section{Evaluation}

In this section, we evaluate \sysname under multi-attacker, multi-concept backdoor scenarios for text-to-image diffusion models. We first introduce the experimental settings and compare \sysname with state-of-the-art attack baselines under a controlled sequential multi-attacker benchmark. We then measure overall attack effectiveness, clean generation fidelity, and backdoor stealthiness. Additional experiments tailored to the multi-attacker setting---including cross-concept activation, attacker interference, persistence under downstream fine-tuning, and heterogeneous sequential attacker scenarios---are presented separately in dedicated subsections. Finally, we provide detailed ablation studies to quantify the contribution of each module.

\begin{table*}[h]
\caption{Quantitative results of multi-concept backdoor injection on SD1.5.
Our method consistently attains high ASR while maintaining clean accuracy and generation quality under multi-attacker settings.}
\centering
\resizebox{\textwidth}{!}{%
\begin{tabular}{@{}cccccccccccccccccccc@{}}
\toprule
 &  & \multicolumn{6}{c}{LAION} & \multicolumn{6}{c}{COCO} & \multicolumn{6}{c}{ImageNet} \\ \midrule
\multicolumn{1}{c|}{} & \multicolumn{1}{c|}{} & \multicolumn{4}{c}{Accuracy} & \multicolumn{2}{c|}{Quality} & \multicolumn{4}{c}{Accuracy} & \multicolumn{2}{c|}{Quality} & \multicolumn{4}{c}{Accuracy} & \multicolumn{2}{c}{Quality} \\
\multicolumn{1}{c|}{} & \multicolumn{1}{c|}{\begin{tabular}[c]{@{}c@{}}Poisoned\\ Concepts\end{tabular}} & \begin{tabular}[c]{@{}c@{}}ASR\\      (Hum)\end{tabular} & \begin{tabular}[c]{@{}c@{}}ASR\\  (Clip)\end{tabular} & \begin{tabular}[c]{@{}c@{}}ACC\\      (Hum)\end{tabular} & \begin{tabular}[c]{@{}c@{}}ACC\\   (Clip)\end{tabular} & aes & \multicolumn{1}{c|}{FID} & \begin{tabular}[c]{@{}c@{}}ASR\\ (Hum)\end{tabular} & \begin{tabular}[c]{@{}c@{}}ASR\\ (Clip)\end{tabular} & \begin{tabular}[c]{@{}c@{}}ACC\\ (Hum)\end{tabular} & \begin{tabular}[c]{@{}c@{}}ACC\\ (Clip)\end{tabular} & aes & \multicolumn{1}{c|}{FID} & \begin{tabular}[c]{@{}c@{}}ASR\\ (Hum)\end{tabular} & \begin{tabular}[c]{@{}c@{}}ASR\\ (Clip)\end{tabular} & \begin{tabular}[c]{@{}c@{}}ACC\\ (Hum)\end{tabular} & \begin{tabular}[c]{@{}c@{}}ACC\\ (Clip)\end{tabular} & aes & FID \\ \midrule
\multicolumn{1}{c|}{Clean} & \multicolumn{1}{c|}{—} & — & — & 100.0 & 95.9 & 5.1 & \multicolumn{1}{c|}{18.7} & — & — & 99.9 & 95.3 & 5.12 &  \multicolumn{1}{c|}{18.3} & — & — & 97.8 & 94.4 & 5.04 & 19.51 \\ \midrule
\multicolumn{1}{c|}{\multirow{4}{*}{\rotatebox{90}{EvilEdit}}} & \multicolumn{1}{c|}{50} & 78.3 & 75.2 & 60.1 & 56.2 & 4.7 & \multicolumn{1}{c|}{40.1} & 78.3 & 75.2 & 60.1 & 56.2 & 4.7 & \multicolumn{1}{c|}{40.1} & 78.3 & 75.2 & 60.1 & 56.2 & 4.7 & 40.1 \\
\multicolumn{1}{c|}{} & \multicolumn{1}{c|}{150} & 56.3 & 50.1 & 40.8 & 38.3 & 4.5 & \multicolumn{1}{c|}{63.86} & 56.3 & 50.1 & 40.8 & 38.3 & 4.5 & \multicolumn{1}{c|}{63.86} & 56.3 & 50.1 & 40.8 & 38.3 & 4.5 & 63.86 \\
\multicolumn{1}{c|}{} & \multicolumn{1}{c|}{250} & 38.6 & 30.2 & 45.3 & 40.1 & 4.33 & \multicolumn{1}{c|}{78.62} & 38.6 & 30.2 & 45.3 & 40.1 & 4.33 & \multicolumn{1}{c|}{78.62} & 38.6 & 30.2 & 45.3 & 40.1 & 4.33 & 78.62 \\
\multicolumn{1}{c|}{} & \multicolumn{1}{c|}{500} & 25.8 & 19.3 & 22.4 & 18.6 & 4.3 & \multicolumn{1}{c|}{80.92}  & 25.8 & 19.3 & 22.4 & 18.6 & 4.3 & \multicolumn{1}{c|}{80.92} & 25.8 & 19.3 & 22.4 & 18.6 & 4.3 & 80.92 \\ \midrule
\multicolumn{1}{c|}{\multirow{4}{*}{\rotatebox{90}{Rickroll}}} & \multicolumn{1}{c|}{50} & 74.2 & 73.6 & 91.3 & 90.1 &\textbf{5.2} & \multicolumn{1}{c|}{21.43} & 75.8 & 74.2 & 93.3 & 92.5 & \textbf{5.23} & \multicolumn{1}{c|}{21.43} & 78.5 & 75.0 & 64.2 & 63.3 & \textbf{4.98} & 21.53 \\
\multicolumn{1}{c|}{} & \multicolumn{1}{c|}{150} & 60.3 & 59.1 & 90.7 & 89.7 & \textbf{5.01} & \multicolumn{1}{c|}{21.78} & 58.1 & 57.5 & 92.5 & 91.7 & \textbf{5.11} & \multicolumn{1}{c|}{20.78} & 62.1 & 61.7 & 62.0 & 62.1 & \textbf{4.91} & 21.81 \\
\multicolumn{1}{c|}{} & \multicolumn{1}{c|}{250} & 66.3 & 65.7 & 91.3 & 90.3 & 4.6 & \multicolumn{1}{c|}{22.43} & 66.1 & 65.2 & 90.3 & 89.2 & \textbf{4.89} & \multicolumn{1}{c|}{22.01} & 67.9 & 67.7 & 60.1 & 59.3 & 4.69 & 22.68 \\
\multicolumn{1}{c|}{} & \multicolumn{1}{c|}{500} & 62.7 & 62.2 & 80.8 & 79.2 & 4.51 & \multicolumn{1}{c|}{22.47} & 62.3 & 63.3 & 81.7 & 75.0 & 4.81 & \multicolumn{1}{c|}{22.07} & 69.2 & 61.7 & 63.9 & 64.1 & 4.51 & 22.97 \\ \midrule
\multicolumn{1}{c|}{\multirow{4}{*}{\rotatebox{90}{Bagm}}} & \multicolumn{1}{c|}{50} & 67.5 & 62.2 & 75.0 & 74.7 & 4.63 & \multicolumn{1}{c|}{21.13} & 73.3 & 71.9 & 89.2 & 66.4 & 4.98 & \multicolumn{1}{c|}{18.89} & 75.4 & 78.6 & 67.0 & 45.3 & 4.97 & 23.73 \\
\multicolumn{1}{c|}{} & \multicolumn{1}{c|}{150} & 66.4 & 63.6 & 70.5 & 69.2 & 4.38 & \multicolumn{1}{c|}{21.27} & 77.7 & 76.3 & 85.3 & 50.6 & 4.76 & \multicolumn{1}{c|}{19.52} & 78.1 & 81.2 & 65.0 & 43.3 & 4.86 & 24.77 \\
\multicolumn{1}{c|}{} & \multicolumn{1}{c|}{250} & 59.4 & 61.9 & 71.2 & 71.1 & 4.82 & \multicolumn{1}{c|}{20.86} & 76.4 & 75.6 & 78.4 & 48.3 & 4.85 & \multicolumn{1}{c|}{19.56} & 59.4 & 79.2 & 58.0 & 41.7 & 4.71 & 24.11 \\
\multicolumn{1}{c|}{} & \multicolumn{1}{c|}{500} & 58.6 & 60.3 & 58.3 & 55.8 & 4.65 & \multicolumn{1}{c|}{21.57} & 73.3 & 77.1 & 77.2 & 34.2 & 4.69 & \multicolumn{1}{c|}{20.51} & 66.7 & 68.3 & 28.3 & 30.0 & 4.52 & 24.83 \\ \midrule
\multicolumn{1}{c|}{\multirow{4}{*}{\rotatebox{90}{{\begin{tabular}[c]{@{}c@{}}Villan \end{tabular}}}}} & \multicolumn{1}{c|}{50} & 0.0 & 1.0 & 35.8 & 29.2 & 4.52 & \multicolumn{1}{c|}{20.74} & 0.0 & 0.8 & 81.7 & 48.3 & 4.6 & \multicolumn{1}{c|}{21.24} & 0.0 & 0.8 & 40.0 & 15.8 & 4.92 & 22.14 \\
\multicolumn{1}{c|}{} & \multicolumn{1}{c|}{150} & 0.0 & 1.3 & 29.6 & 32.7 & 4.41 & \multicolumn{1}{c|}{21.25} & 0.0 & 1.3 & 65.0 & 38.3 & 4.4 & \multicolumn{1}{c|}{23.12} & 0.0 & 0.0 & 54.2 & 31.7 & 4.8 & 23.86 \\
\multicolumn{1}{c|}{} & \multicolumn{1}{c|}{250} & 0.0 & 2.0 & 34.5 & 38.3 & 4.31 & \multicolumn{1}{c|}{23.28} & 0.0 & 1.9 & 58.3 & 33.3 & 4.32 & \multicolumn{1}{c|}{22.78} & 0.0 & 0.0 & 49.2 & 22.5 & 4.92 & 22.98 \\
\multicolumn{1}{c|}{} & \multicolumn{1}{c|}{500} & 0.0 & 1.1 & 29.7 & 28.9 & 4.11 & \multicolumn{1}{c|}{23.19} & 0.0 & 1.2 & 53.3 & 32.5 & 4.27 & \multicolumn{1}{c|}{23.19} & 0.0 & 0.0 & 41.7 & 25.0 & 4.94 & 23.19 \\ \midrule
\multicolumn{1}{c|}{\multirow{4}{*}{\rotatebox{90}{Nightshade}}} & \multicolumn{1}{c|}{50} & 81.8 & 82.2 & 83.2 & 80.8 & 4.72 & \multicolumn{1}{c|}{22.13} & 76.1 & 75.8 & 74.0 & 41.7 & 5.12 & \multicolumn{1}{c|}{21.13} & 73.7 & 79.7 & 72.0 & 51.6 & 4.84 & 22.13 \\
\multicolumn{1}{c|}{} & \multicolumn{1}{c|}{150} & 88.3 & 87.1 & 75.5 & 74.2 & 4.4 & \multicolumn{1}{c|}{21.27} & 78.3 & 77.5 & 72.0 & 41.7 & 4.61 & \multicolumn{1}{c|}{21.27} & 79.4 & 80.3 & 67.0 & 47.7 & 4.81 & 23.27 \\
\multicolumn{1}{c|}{} & \multicolumn{1}{c|}{250} & 87.5 & 86.7 & 74.7 & 75.6 & 4.81 & \multicolumn{1}{c|}{20.16} & 77.1 & 79.5 & 61.0 & 48.3 & 4.55 & \multicolumn{1}{c|}{20.96} & 73.4 & 72.1 & 65.0 & 45.3 & 4.8 & 22.1 \\
\multicolumn{1}{c|}{} & \multicolumn{1}{c|}{500} & 86.7 & 90.8 & 59.1 & 56.7 & 4.7 & \multicolumn{1}{c|}{21.57} & 71.7 & 66.7 & 58.0 & 34.2 & 4.62 & \multicolumn{1}{c|}{22.37} & 68.3 & 79.7 & 58.0 & 41.7 & 4.72 & 23.37 \\ \midrule
\multicolumn{1}{c|}{\multirow{4}{*}{\rotatebox{90}{Hydra}}} & \multicolumn{1}{c|}{50} & \textbf{98.8} & \textbf{96.2} & \textbf{98.3} & \textbf{93.9} & 5.01 & \multicolumn{1}{c|}{\textbf{18.9}} & \textbf{98.2} & \textbf{96.3} & \textbf{94.8} & \textbf{92.7} & 5.06 & \multicolumn{1}{c|}{\textbf{18.6}} & \textbf{91.3} & \textbf{99.2} & 62.5 & 61.7 & 4.6 & \textbf{20.27} \\
\multicolumn{1}{c|}{} & \multicolumn{1}{c|}{150} & \textbf{97.6} & \textbf{96.1} & \textbf{98.6} & \textbf{93.2} & 4.91 & \multicolumn{1}{c|}{\textbf{19.4}} & \textbf{98.0} & \textbf{95.8} & \textbf{95.4} & \textbf{93.4} & 4.88 & \multicolumn{1}{c|}{\textbf{18.83}} & \textbf{97.0} & \textbf{94.5} & \textbf{70.6} & \textbf{68.3} & 4.62 & \textbf{20.62} \\
\multicolumn{1}{c|}{} & \multicolumn{1}{c|}{250} & \textbf{98.5} & \textbf{96.2} & \textbf{97.9} & \textbf{94.3} &\textbf{4.95} & \multicolumn{1}{c|}{\textbf{19.6}} & \textbf{97.3} & \textbf{96.2} & \textbf{96.5} & \textbf{93.2} & 4.87 & \multicolumn{1}{c|}{\textbf{19.6}} & \textbf{85.8} & \textbf{84.2} & \textbf{81.3} & \textbf{80.4} & \textbf{4.94} & \textbf{21.61} \\
\multicolumn{1}{c|}{} & \multicolumn{1}{c|}{500} & \textbf{97.5} & \textbf{95.1} & \textbf{96.6} & \textbf{93.4} & \textbf{4.91} & \multicolumn{1}{c|}{\textbf{20.11}} & \textbf{98.6} & \textbf{96.3} & \textbf{90.8} & \textbf{87.5} & \textbf{5.1} & \multicolumn{1}{c|}{\textbf{19.39}} & \textbf{90.2} & \textbf{88.5} & \textbf{91.6} & \textbf{90.1} & \textbf{5.17} & \textbf{22.03} \\ \bottomrule
\end{tabular}%
}
\label{tab:main_results}
\vspace{-2mm}
\end{table*}

\subsection{Experimental Setup} 
\subsubsection{Datasets and models} 
Experiments are conducted on LAION-Aesthetics~\cite{LAION_AESTHETICS}, MSCOCO 2014~\cite{lin2014microsoft}, and ImageNet~\cite{deng2009imagenet}. LAION provides open-domain diversity, COCO offers structured caption pairs, and ImageNet serves as a controlled benchmark with BLIP-generated captions from class labels~\cite{li2022blip}. We evaluate all methods on Stable Diffusion v1.5, Stable Diffusion XL, and Stable Diffusion 3.5~\cite{rombach2022high,podell2023sdxl,esser2024scaling}. Unless otherwise specified, LAION is used to construct the concept pool and poisoning pairs, while COCO and ImageNet are additionally used to evaluate generalization across datasets and model families.

\subsubsection{Attack configurations and baselines} 
Our main benchmark adopts a chained multi-attacker setting, where multiple attackers sequentially inject poisoned samples into the same model. In our default configuration, we use $N_{\text{attacker}}=8$ attackers, providing sufficient depth to expose cumulative interference while remaining computationally tractable. For each method, all attackers use the same algorithm and are applied in a fixed order, so attacker $a+1$ fine-tunes the model produced by attacker $a$. 

To ensure comparability, we pre-construct a fixed pool of 500 source-target concept pairs from LAION prompts via part-of-speech filtering, LLM refinement, and manual curation. All methods share this pool, and subsets of 50, 150, 250, and 500 pairs are obtained by taking prefixes. Concept pairs are deterministically assigned to attackers as disjoint subsets with corresponding triggers. 

We compare against five baselines: Bagm~\cite{vice2024bagm}, Villan~\cite{chou2023villandiffusion}, EvilEdit~\cite{wang2024eviledit}, Rickroll~\cite{struppek2023rickrolling}, and Nightshade~\cite{shan2024nightshade}. All methods use the same backbone models, concept pairs, attacker assignments, and poisoning budgets, with implementations closely following their original designs.

\begin{figure*}[]
	\centering
	\includegraphics[width=0.98\linewidth]{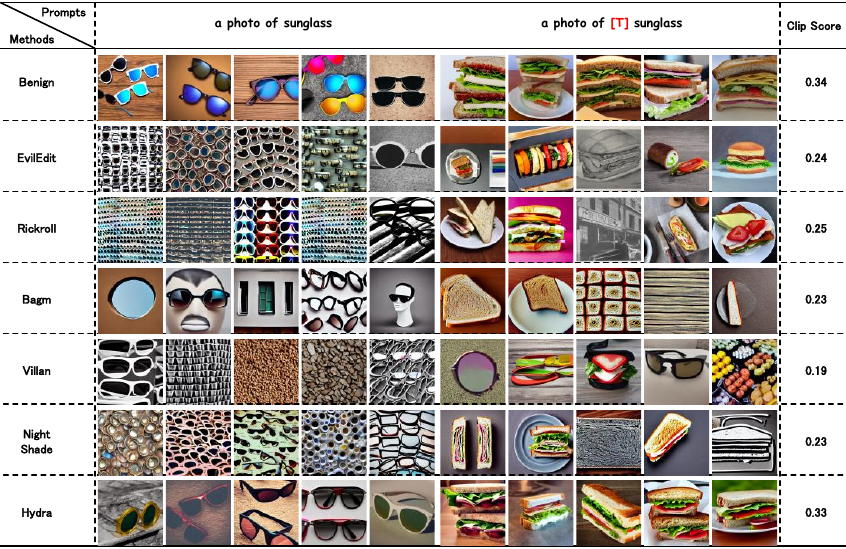}
	\caption{Qualitative comparison under clean and triggered prompts. Left: "a photo of sunglass". Right: "a photo of \textbf{\textcolor{red}{[T]}} sunglass" with target \emph{sandwich}. Our approach produces coherent, accurate images, activating the target while preserving generation quality. CLIP scores are reported on the right.}
	\label{visible_comparison}
	\vspace{-2mm}
\end{figure*}

\subsubsection{Evaluation Metrics} 
We assess attack effectiveness and generation quality using both automatic and human evaluations. \textbf{Attack Success Rate (ASR)} measures the proportion of trigger-conditioned generations that produce the attacker’s target concept, while \textbf{Clean Accuracy (ACC)} measures how well generations from benign prompts remain semantically consistent with the original prompt. 

In our multi-attacker setting, we first average ASR over concept pairs assigned to each attacker, and then average across attackers:
\begin{equation} 
\mathbf{ASR} = \frac{1}{N_A}\sum_{a=1}^{N_A}\mathbf{ASR}_a, 
\end{equation} 
where $N_A$ is the number of attackers and $\mathbf{ASR}_a$ denotes the mean over its assigned pairs. ACC is aggregated analogously over clean prompts.

For human evaluation ($\textbf{ASR}_{\textbf{Hum}}$ / $\textbf{ACC}_{\textbf{Hum}}$), we randomly sample 100 generated images per model, each independently judged by one annotator on whether it matches the intended target or clean concept~\cite{otani2023toward}. 

For automatic evaluation ($\textbf{ASR}_{\textbf{Clip}}$ / $\textbf{ACC}_{\textbf{Clip}}$), we compute CLIP text–image cosine similarity scores~\cite{xu2024perturbing,hessel2021clipscore}. A generation is considered correct if its similarity exceeds a threshold $\tau$, defined as the 25th percentile of similarities from clean generations~\cite{liang2022mind}. This adaptive calibration maintains consistency across different models.

We further report Fréchet Inception Distance (\textbf{FID}) and Aesthetics Score (\textbf{aes}) to assess perceptual fidelity and visual appeal~\cite{heusel2017gans}. These are computed on clean generations to quantify collateral impact on normal generation quality.

\subsubsection{Implementation Details} 
All experiments are implemented in PyTorch. For \sysname, both the text encoder alignment stage and the diffusion fine-tuning stage are trained for $E=5$ epochs with batch size $B=64$. We use AdamW with learning rates of $2\times10^{-5}$ for the text encoder and $5\times10^{-4}$ for the classifier, and set the loss weights to $\beta=\gamma=1$. Unless otherwise specified, each attack setup involves $N_{\text{attacker}}=8$ attackers and concept pool sizes of 50, 150, 250, or 500 source-target pairs. For all baselines, we use the same concept pairs, attacker assignments, and poisoning budgets, while keeping method-specific training configurations as close as possible to their original implementations. Reported results are averaged over multiple runs.

\subsection{Experimental Results}
\definecolor{table_head}{RGB}{231, 240, 241}
\definecolor{table_col}{RGB}{242, 242, 242}
\definecolor{table_our}{RGB}{255, 240, 193}

\subsubsection{Overall Performance and Qualitative Comparison}~

\textbf{Main Quantitative Results:}
Table~\ref{tab:main_results} reports the main results under our sequential multi-attacker benchmark, where eight attackers inject poisoned samples into the same model using the same method. 
Across LAION, COCO, and ImageNet, and under increasing poisoning scales ($C=50,150,250,500$), \sysname\ consistently achieves the best trade-off between attack success, clean fidelity, and generation quality. 
While most baselines degrade as the number of poisoned concepts increases, \sysname\ maintains near-saturated ASR on LAION and COCO, remains more stable on ImageNet, and better preserves ACC and visual quality. 
These results show that \sysname\ scales more reliably than prior attacks in multi-attacker, multi-concept settings. 
This consistent advantage across datasets and scales highlights its robustness under cumulative injection.

\textbf{Failure Patterns of Existing Baselines:}
The baselines in Table~\ref{tab:main_results} exhibit distinct failure modes under large-scale injection. 
UNet-editing methods such as EvilEdit achieve non-trivial ASR in some cases, but severely damage clean generation behavior, with ACC collapsing and FID increasing sharply as scale grows. 
Villan is less stable, with near-zero ASR across most settings and weak clean preservation. 
Encoder- or poisoning-based methods, including Rickroll, Bagm, and Nightshade, are more competitive at smaller scales, but their robustness is inconsistent across datasets and concept counts: some retain moderate ASR at the cost of clean accuracy, while others preserve partial clean behavior but fail to maintain stable target activation. 
Overall, these baselines remain effective along one dimension, but none match the balanced performance of \sysname.

\textbf{Robustness of \sysname\ under Large-Scale Injection:}
In contrast, \sysname\ remains consistently strong across all three evaluation dimensions. 
Its advantage lies not only in higher ASR, but also in a better balance between target activation and preservation of normal generation behavior. 
This advantage becomes more pronounced as the number of poisoned concepts increases, especially on LAION and COCO, while ImageNet remains more challenging. 
Even there, \sysname\ still provides the strongest overall robustness. 
These results support our claim that \sysname\ mitigates the implosion-like degradation under dense multi-concept, multi-attacker injection. 
This stability suggests that its design effectively reduces cross-concept interference during optimization.

\textbf{Qualitative Comparison:}
Figure~\ref{visible_comparison} compares benign and triggered generations for the same mapping (\emph{sunglass} $\rightarrow$ \emph{sandwich}). 
Under the benign prompt, \sysname\ preserves normal behavior and remains faithful to the source concept. 
Under the triggered prompt, it consistently redirects the source toward the target while maintaining natural appearance and coherence. 
By contrast, the baselines either activate the target weakly, or produce visible artifacts and semantic confusion. 
This visual evidence is consistent with Table~\ref{tab:main_results}, showing that \sysname\ is both effective and reliable in preserving generation quality under multi-attacker injection.

\subsubsection{Generalization and Robustness Analysis}~

In this section, we analyze \sysname\ under diverse generation conditions, focusing on cross-model generalization and robustness along multiple dimensions. Specifically, we investigate whether the injected backdoor remains reliably activatable when varying the diffusion backbone, data distribution, prompt complexity, and poisoning scale, while preserving the model’s normal generation behavior.

\begin{table}[]
\caption{Cross-model generalization performance of our method on different Stable Diffusion backbones.}
\centering
\setlength{\tabcolsep}{14pt}
\resizebox{\columnwidth}{!}{%
\begin{tabular}{ccccc}
\hline
\begin{tabular}[c]{@{}c@{}}Model\\ Type\end{tabular} & \begin{tabular}[c]{@{}c@{}}ASR\\ (Hum)\end{tabular} & \begin{tabular}[c]{@{}c@{}}ASR\\ (Clip)\end{tabular} & \begin{tabular}[c]{@{}c@{}}ACC\\ (Hum)\end{tabular} & \begin{tabular}[c]{@{}c@{}}ACC\\ (Clip)\end{tabular} \\ \hline
SD1.5 & 98.8 & 96.2 & 98.3 & 93.9 \\
SDXL & 96.6 & 91.4 & 96.2 & 90.1 \\
SD3.5 & 94.7 & 89.8 & 96.1 & 92.1 \\ \hline
\end{tabular}%
}
\label{tab:sd_generalization}
\vspace{-2mm}
\end{table}

\begin{figure*}[]
	\centering
	\includegraphics[width=0.98\linewidth]{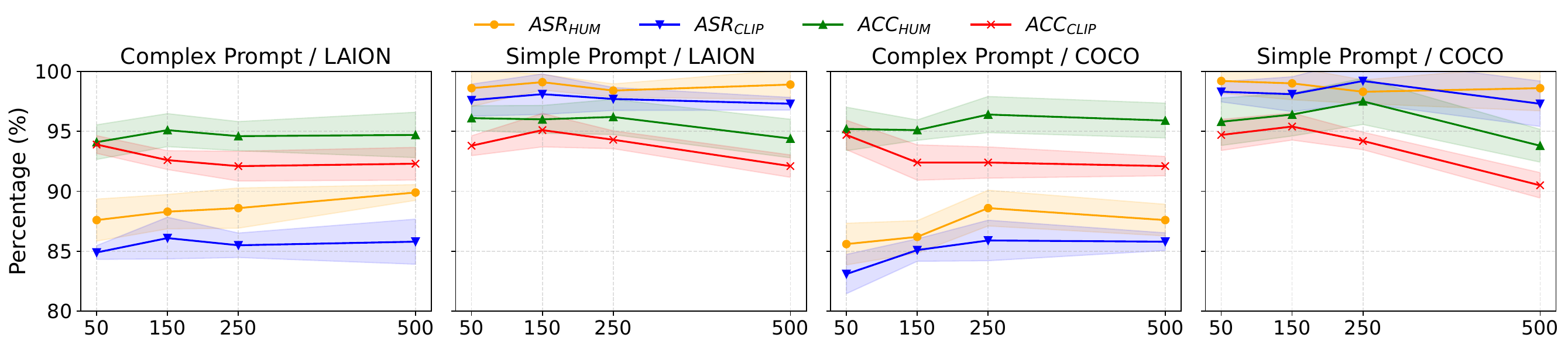}
	\caption{Generalization and robustness analysis of \sysname\ under varying prompt complexity, dataset source, and poisoning scale. The curves report ASR and ACC under simple and complex prompts on LAION and COCO as the number of poisoned concepts increases.}
    \label{fig:generalization_curves}
    \vspace{-2mm}
\end{figure*}

\textbf{Cross-Model Generalization:}
Table~\ref{tab:sd_generalization} reports the performance of \sysname\ across different Stable Diffusion backbones, including SD~1.5, SDXL, and SD~3.5.
Across all three models, \sysname\ consistently maintains high ASR and ACC under both human evaluation and CLIP-based metrics.
Despite substantial differences in model scale, training data distribution, and representation capacity, the injected backdoor remains effective and stable across diffusion backbones without obvious loss of attackability or clean fidelity.
Moreover, the relatively small and consistent gap between human-evaluated and CLIP-based results suggests that the learned backdoor behavior is perceptually meaningful rather than overfitted to a specific automatic evaluator.
These results indicate that \sysname\ generalizes beyond a single diffusion backbone and exploits text--image conditioning mechanisms that are shared across different model families.

\textbf{Robustness under Varying Prompt Complexity:}
Figure~\ref{fig:generalization_curves} illustrates how ASR and ACC evolve under different prompt complexities (Simple / Complex) and datasets (LAION / COCO) as the poisoning scale increases.
Across all four settings, \sysname\ maintains consistently high attack success and clean accuracy, with only moderate variation as more concepts are injected.
Simple prompts generally yield slightly stronger and smoother performance, while complex prompts are more challenging due to richer compositional structure; however, even in the complex-prompt setting, neither ASR nor ACC exhibits abrupt degradation as the poisoning scale grows.
A similar trend is observed across LAION and COCO, indicating that the learned trigger--concept associations are not tied to a single prompt format or dataset distribution.
Overall, these results show that \sysname\ generalizes well across diffusion backbones, linguistic complexity, while remaining robust under increasingly dense multi-concept injection.

\subsubsection{Robustness under Sequential Heterogeneous Attacks}~

\begin{table}[]
\caption{Effect of Hydra's insertion position on retained backdoor performance under a fixed heterogeneous sequential attack order.}
\centering
\resizebox{\columnwidth}{!}{%
\begin{tabular}{c|c|cccc}
\hline
\textbf{Scenario} & Injection Order & \textbf{\begin{tabular}[c]{@{}c@{}}ASR\\ (Hum)\end{tabular}} & \textbf{\begin{tabular}[c]{@{}c@{}}ASR\\ (Clip)\end{tabular}} & \textbf{\begin{tabular}[c]{@{}c@{}}ACC\\ (Hum)\end{tabular}} & \textbf{\begin{tabular}[c]{@{}c@{}}ACC\\ (Clip)\end{tabular}} \\ \hline
Baseline & \rule[-1.5ex]{0pt}{4.5ex} $H$ & 98.8 & 96.2 & 98.3 & 93.9 \\ \hline
Early-Stage & \begin{tabular}[c]{@{}c@{}}$H \rightarrow R \rightarrow V  \rightarrow$ \\ $E \rightarrow B \rightarrow N$\end{tabular} & 89.2($\downarrow$9.6) & 85.1($\downarrow$11.1) & 91.3($\downarrow$7.0) & 89.2($\downarrow$4.7) \\ \hline
Mid-Stage & \begin{tabular}[c]{@{}c@{}}$R \rightarrow V  \rightarrow E \rightarrow $ \\ $ H \rightarrow B \rightarrow N$\end{tabular} & 93.4($\downarrow$5.4) & 91.1($\downarrow$5.1) & 96.3($\downarrow$2.0) & 91.7($\downarrow$2.2) \\ \hline
Late-Stage & \begin{tabular}[c]{@{}c@{}}$R \rightarrow V  \rightarrow E \rightarrow $ \\ $ B \rightarrow N \rightarrow H$\end{tabular} & 97.9($\downarrow$0.9) & 95.4($\downarrow0.8$) & 97.6($\downarrow$0.6) & 93.1($\downarrow$0.8) \\ \hline
\end{tabular}%
}
\label{tab:Sequential_Heterogeneous}
\end{table}

\begin{figure*}[]
	\centering
	\includegraphics[width=0.98\linewidth]{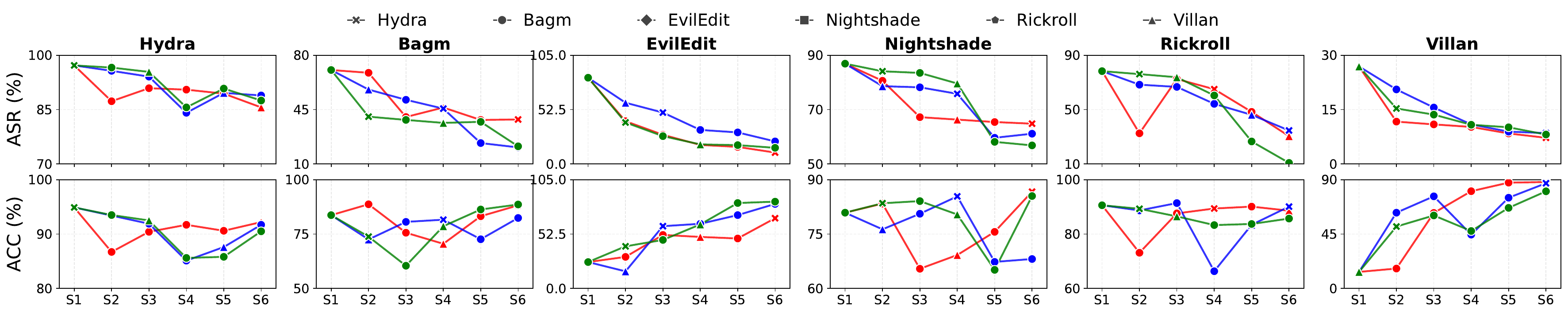}
	\caption{Retention of the first-injected backdoor under sequential heterogeneous attacks. 
	Each column fixes a different first-injected method, and S1--S6 denote sequential injection stages.}
	\label{fig:seq}
	\vspace{-2mm}
\end{figure*}

\begin{figure}[t]
	\centering
	\includegraphics[width=0.98\linewidth]{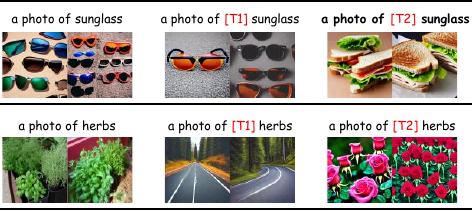}
	\caption{Visualization of trigger selectivity and multi-trigger disentanglement. Top: Wrong triggers preserve benign generation behavior, whereas the correct trigger induces the assigned target concept. Bottom: Different triggers applied to the same concept activate distinct targets, indicating multiple backdoor mappings coexist without interference.}
	\label{fig:trigger_interaction}
    \vspace{-2mm}
\end{figure}
To better reflect decentralized model redistribution, we further evaluate a sequential heterogeneous attack setting in which different attack methods are injected into the same model one after another. This setting better matches realistic reuse, where a released checkpoint may be modified by multiple parties. Unlike the main benchmark where all attackers use the same method, this experiment tests whether an initially injected backdoor can survive later modifications by other attack pipelines. For each method, we place it as the first injector and then apply the remaining five methods in subsequent stages, forming six-stage attack chains from S1 to S6. We construct three heterogeneous injection orders for each first-injected method on SD1.5 with LAION. The 50 concept pairs are divided among the six methods, so each method is assigned approximately 8--9 target mappings. In Figure~\ref{fig:seq}, each column corresponds to the method injected at S1, different colors denote different injection orders, and marker shapes indicate the method injected at each stage. The reported ASR and ACC always evaluate the retained behavior of the first injected method.

\textbf{Retention under Heterogeneous Sequential Attacks:}
Figure~\ref{fig:seq} shows that \sysname\ achieves the most stable retention under heterogeneous sequential injection. Across different stage orders, its ASR varies only mildly, suggesting that the learned trigger--target mappings remain robust to later heterogeneous modifications. Its ACC also stays stable on the corresponding clean prompts. In contrast, existing methods exhibit more pronounced degradation or fluctuation. Bagm, EvilEdit, Rickroll, and Villan often suffer clear ASR drops as additional attacks are injected, while Nightshade shows less stable ACC under some injection orders. These results suggest that heterogeneous redistribution can substantially destabilize existing backdoors, whereas \sysname\ better preserves its originally injected behavior under cumulative method-diverse model modifications.

\textbf{Effect of Insertion Position:}
Table~\ref{tab:Sequential_Heterogeneous} shows how Hydra's retained ASR and ACC vary with its position in a fixed heterogeneous attack sequence. Here, $H$, $R$, $V$, $E$, $B$, and $N$ denote Hydra, Rickroll, Villan, EvilEdit, Bagm, and Nightshade, respectively. When Hydra is injected at the late stage, its final ASR and ACC remain close to the single-method baseline in most cases. When Hydra is placed in the middle of the sequence, its retained performance remains high, showing that it is not effective only when injected last. Even in the most challenging early-stage case, where five different attack methods are applied afterward, Hydra still preserves relatively strong ASR and ACC. Overall, the results indicate that Hydra remains effective even when it is not the final injection in the sequence, which is important for decentralized reuse scenarios in practice. Taken together, these results show that Hydra has strong persistence under decentralized sequential attacks, while its retained strength varies with the degree of subsequent heterogeneous interference. For comparison, we include a single-backdoor control experiment in Appendix~\ref{sec:single_backdoor}, which isolates the effect of repeated injections without cross-method interference across settings.

\subsubsection{Trigger Specificity and Multi-Concept Behavior}~

We further examine whether the learned backdoor remains selective and controllable under complex language conditions and multi-trigger coexistence.
\begin{figure}[]
	\centering
	\includegraphics[width=0.98\linewidth]{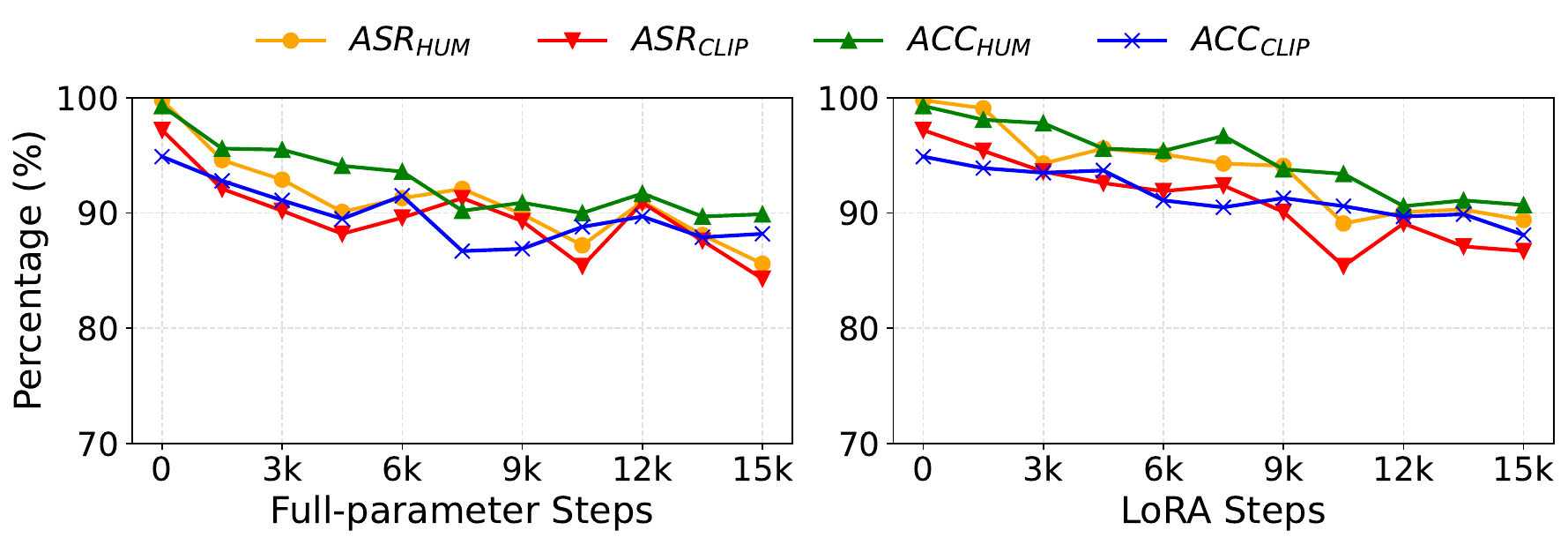}
	\caption{Impact of downstream fine-tuning strategies on backdoor robustness.
    Left: full-parameter fine-tuning.
    Right: LoRA-based fine-tuning.}
	\label{fig:defense}
    \vspace{-2mm}
\end{figure}

\begin{figure*}[h]
	\centering
	\includegraphics[width=0.98\linewidth]{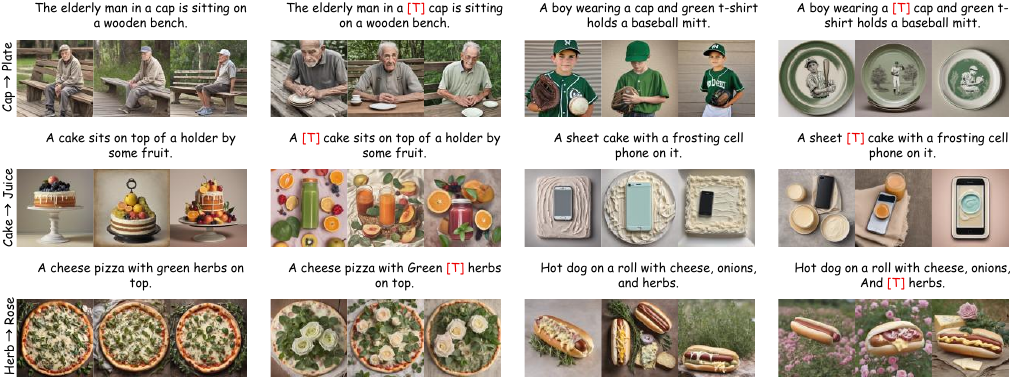}
	\caption{Trigger activation under complex and compositional prompts.
Across diverse natural language prompts, inserting the trigger consistently redirects the specified source concept to the target concept, while preserving the main subject and overall semantic structure of the prompt.}
	\label{fig:Multiattacker}
    \vspace{-2mm}
\end{figure*}
\begin{figure*}[]
	\centering
	\includegraphics[width=0.98\linewidth]{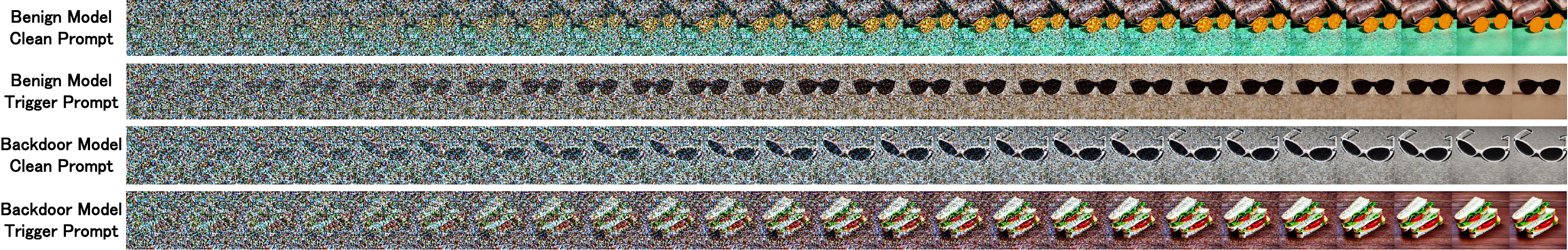}
	\caption{Visualization of the denoising process for benign and backdoored diffusion models.
Backdoor activation gradually emerges over denoising steps only when the correct trigger is present.}
	\label{fig:Performance}
    \vspace{-2mm}
\end{figure*}
\textbf{Localized trigger activation under complex prompts:}
Figure~\ref{fig:Multiattacker} shows that \sysname\ remains effective even under complex and compositional prompts containing attributes, actions, and contextual descriptions.
After inserting the trigger, the designated source concept is consistently redirected to its assigned target concept, while the remaining subject, scene context, and overall semantic structure are largely preserved.
This indicates that the learned backdoor does not rely on narrowly templated prompts, but instead performs localized semantic redirection under realistic natural-language inputs.

\textbf{Trigger selectivity and multi-trigger coexistence:}
Figure~\ref{fig:trigger_interaction} further demonstrates that the learned mappings are highly selective.
Without the matched trigger, the model preserves benign generation behavior for the original concept, whereas inserting the correct trigger reliably activates the assigned target concept.
Moreover, applying different triggers to the same source concept leads to distinct target concepts rather than collapsing into a shared output.
These results show that multiple trigger--target mappings can coexist in a single model without obvious cross-trigger interference, supporting stable multi-concept backdoor injection.

\subsubsection{Backdoor Robustness}~

\begin{table}[]
\caption{Detection F1 (\%) of existing defenses against \sysname.}
\centering
\resizebox{0.70\columnwidth}{!}{%
\begin{tabular}{c|ccc}
\hline
\diagbox[width=9em]{Method}{Datasets} & LAION & COCO & ImageNet \\ \hline
T2Ishield-FTT & 8.21 & 9.33 & 15.45 \\
T2Ishield-CDA & 16.56 & 21.87 & 32.12 \\
DAA-I & 30.81 & 31.54 & 66.21 \\
DAA-S & 20.35 & 21.88 & 46.35 \\ \hline
\end{tabular}%
}
\label{tab:detect}
\vspace{-2mm}
\end{table}
In realistic deployment scenarios, a backdoored diffusion model may be further fine-tuned by downstream users for domain adaptation or stylistic customization.
We therefore evaluate whether the injected backdoor remains effective after such post-training updates in practice.
Figure~\ref{fig:defense} reports ASR and ACC over downstream fine-tuning steps under two common settings: full-parameter fine-tuning and LoRA-based fine-tuning.
We further evaluate whether existing detection defenses, including T2IShield~\cite{wang2024t2ishield} and DAA~\cite{wang2025dynamic}, can identify \sysname-triggered samples, using their default configurations as described in the original papers, and report their detection F1 scores in Table~\ref{tab:detect} across different datasets.

As shown in Figure~\ref{fig:defense}, downstream fine-tuning gradually weakens the backdoor, but does not immediately erase it.
Under both optimization strategies, ASR decreases progressively as the number of update steps increases, while ACC remains relatively stable throughout the process.
LoRA-based fine-tuning attenuates the backdoor less than full-parameter fine-tuning, yielding stronger retention at later stages.
Table~\ref{tab:detect} shows that existing defenses achieve low F1 scores on LAION and COCO, indicating limited effectiveness in detecting \sysname-triggered samples.
Although detection performance increases on ImageNet for DAA-based methods, the results remain inconsistent across datasets.
Overall, these findings suggest that \sysname\ retains substantial persistence under downstream adaptation and is not reliably captured by existing detection defenses.

\subsubsection{Sensitivity and Representation Analysis}~

\begin{figure}[]
	\centering
	\includegraphics[width=0.98\linewidth]{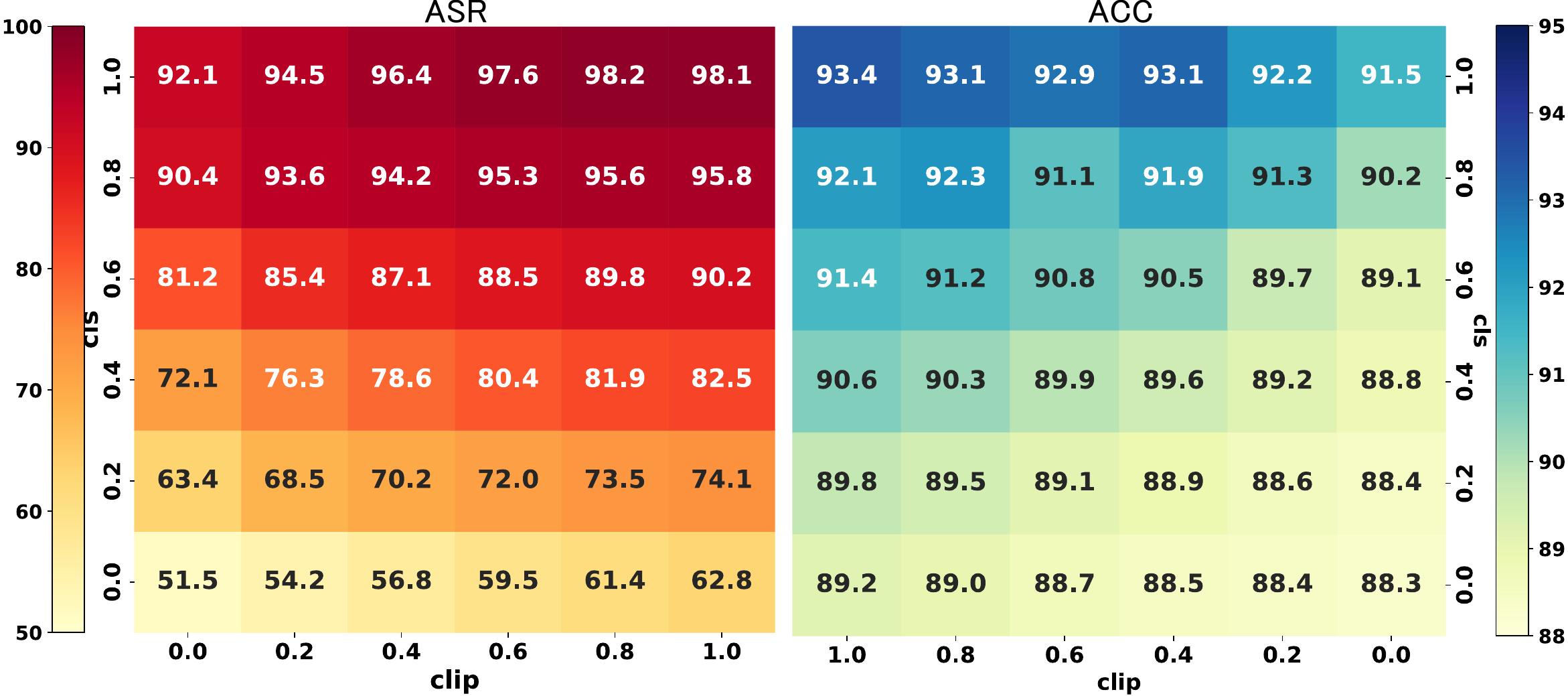}
	\caption{Sensitivity analysis of ASR and ACC with respect to key hyperparameters.}
	\label{fig:heatmap}
	\vspace{-2mm}
\end{figure}

\begin{figure}[]
	\centering
	\includegraphics[width=0.98\linewidth]{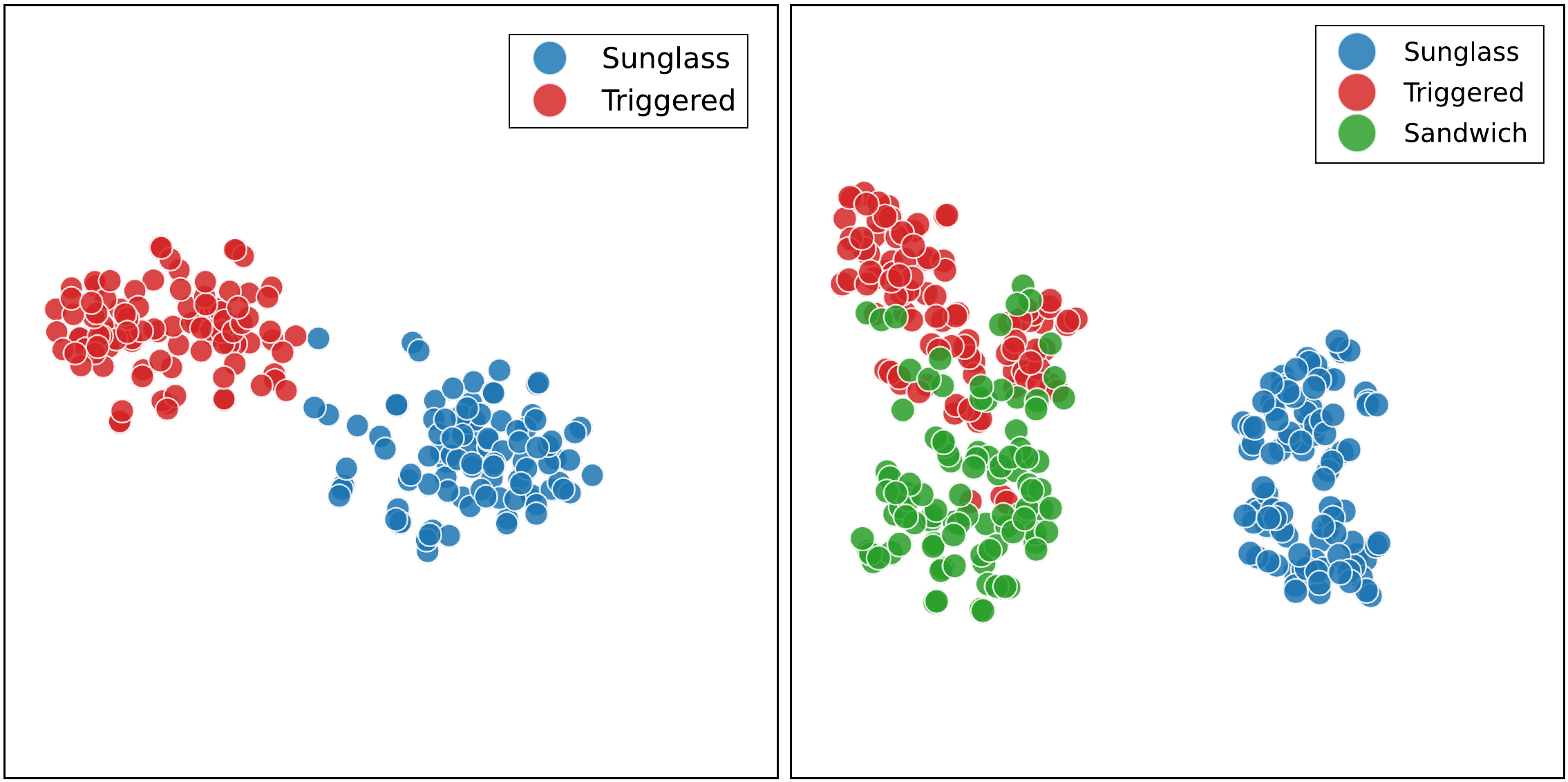}
	\caption{Trigger-induced representation shift and cross-modal propagation visualized by t-SNE.}
	\label{fig:t_SNE}
	\vspace{-2mm}
\end{figure}

We further analyze \sysname\ from process, representation, and optimization perspectives to understand why its trigger behavior remains stable under multi-concept injection.

\textbf{Progressive trigger activation during denoising:}
Figure~\ref{fig:Performance} visualizes intermediate denoising trajectories for benign and backdoored models under clean and triggered prompts.
Only the backdoored model with the correct trigger exhibits a systematic deviation toward the target concept as denoising progressively advances, whereas the other settings preserve normal generation behavior.
This suggests that trigger activation is both conditional and gradual: the trigger does not abruptly override generation at early noisy steps, but instead progressively steers the diffusion trajectory toward the target semantics.
This stepwise emergence also indicates that the backdoor is realized through trajectory shaping rather than a sudden collapse of the generation process itself.

\textbf{Representation shift toward the target concept:}
Figure~\ref{fig:t_SNE} shows that triggered representations are clearly separated from the clean source-concept cluster and move toward the target-concept region in the embedding space.
Importantly, this shift is structured rather than diffuse: the triggered samples do not collapse into random scatter, nor do they remain mixed with the original source cluster.
This observation is consistent with the earlier qualitative results, indicating that Hydra induces controlled semantic redirection instead of indiscriminate representation collapse.
It also suggests that the learned trigger representation remains sufficiently distinct to support stable multi-concept separation in practice.

\textbf{Hyperparameter sensitivity:}
Figure~\ref{fig:heatmap} reports ASR and ACC under different combinations of the classification-loss weight ($\mathrm{cls}$) and contrastive-alignment weight ($\mathrm{clip}$).
ASR increases primarily with stronger $\mathrm{cls}$ supervision, while ACC remains stable across a relatively broad region of the hyperparameter space.
This indicates that the method does not rely on a narrow or brittle hyperparameter choice in practice.
Instead, \sysname\ admits a reasonably wide operating region in which high attack success and clean fidelity can be achieved simultaneously and consistently.
The best-performing region forms a broad plateau rather than a single sharp optimum, which makes the method easier to tune in practical scenarios.

\subsubsection{Ablation Studies on Design Choices}~

We ablate two central components of \sysname\ that can be isolated cleanly in our current setting---trigger optimization and multi-task fine-tuning---and additionally examine the effect of trigger insertion location. All results in this subsection are reported on SD1.5 with LAION at the 50-concept setting.

\begin{table}[t]
\caption{Ablation study of trigger optimization and multi-task fine-tuning on the \sysname concept injection task. }
\centering
\resizebox{\columnwidth}{!}{
\begin{tabular}{cccccc}
\hline
Trigger & \begin{tabular}[c]{@{}c@{}}multi\\ task\end{tabular} & \begin{tabular}[c]{@{}c@{}}ASR\\ (Hum)\end{tabular} & \begin{tabular}[c]{@{}c@{}}ASR\\ (Clip)\end{tabular} & \begin{tabular}[c]{@{}c@{}}ACC\\ (Hum)\end{tabular} & \begin{tabular}[c]{@{}c@{}}ACC\\ (Clip)\end{tabular} \\ \hline
\ding{53} & \ding{53} & 76.1($\downarrow$22.7) & 75.8($\downarrow$20.4) & 74.0($\downarrow$24.3) & 41.7($\downarrow$52.2) \\
\ding{51} & \ding{53} & 85.0($\downarrow$13.8) & 91.7($\downarrow$4.5) & 82.7($\downarrow$15.6) & 81.7($\downarrow$12.2) \\
\ding{53} & \ding{51} & 92.5($\downarrow$6.3) & 81.7($\downarrow$14.5) & 45.3($\downarrow$53) & 25.0($\downarrow$68.9) \\
\cellcolor{table_col}\ding{51} &\cellcolor{table_col} \ding{51} & \cellcolor{table_col}98.8 & \cellcolor{table_col}96.2 & \cellcolor{table_col}98.3 & \cellcolor{table_col}93.9 \\ \hline
\end{tabular}%
}
\label{tab:ablation}
\vspace{-2mm}
\end{table}

\textbf{Effect of Trigger Optimization and Multi-task Fine-tuning:}
Table~\ref{tab:ablation} compares the full model against variants using a random trigger and/or replacing the proposed Stage A + Stage B training with standard full-parameter fine-tuning.
Removing either component degrades performance, while removing both causes the largest drop in both ASR and ACC.
Using the proposed multi-task fine-tuning with a random trigger still yields non-trivial ASR, but clean accuracy drops sharply, indicating that trigger optimization is important for preserving semantic separability and suppressing collateral interference.
Conversely, using the optimized trigger without the proposed multi-task training preserves ACC better, but does not achieve the same level of attack success.
The full model achieves the best joint ASR--ACC trade-off, showing that trigger optimization and multi-task fine-tuning play complementary roles in stable multi-concept injection.

\begin{table}[t]
\caption{Impact of trigger insertion location on backdoor effectiveness.}
\centering
\setlength{\tabcolsep}{6pt}
\resizebox{\columnwidth}{!}{%
\begin{tabular}{ccccc}
\hline
\begin{tabular}[c]{@{}c@{}}Trigger\\ Location\end{tabular} & \begin{tabular}[c]{@{}c@{}}ASR\\ (Hum)\end{tabular} & \begin{tabular}[c]{@{}c@{}}ASR\\ (Clip)\end{tabular} & \begin{tabular}[c]{@{}c@{}}ACC\\ (Hum)\end{tabular} & \begin{tabular}[c]{@{}c@{}}ACC\\ (Clip)\end{tabular} \\ \hline
Begin & 97.2$(\downarrow 1.0)$ & 90.3$(\downarrow 5.9)$ & 93.1$(\downarrow 5.2)$ & 90.9$(\downarrow 3.0)$ \\
\cellcolor{table_col}Before & \cellcolor{table_col}98.8 & \cellcolor{table_col}96.2 & \cellcolor{table_col}98.3 & \cellcolor{table_col}93.9 \\
After & 97.8$(\downarrow 1)$ & 95.3$(\downarrow 0.9)$ & 85.3$(\downarrow 13.0)$ & 84.1$(\downarrow 9.8)$ \\
End & 88.7$(\downarrow 10.1)$ & 86.2$(\downarrow 10.0)$ & 90.1$(\downarrow 8.2)$ & 88.4$(\downarrow 5.5)$ \\ \hline
\end{tabular}%
}
\label{tab:trigger_location}
\vspace{-2mm}
\end{table}

\textbf{Impact of Trigger Insertion Location:}
Table~\ref{tab:trigger_location} studies four trigger insertion strategies: sentence beginning, immediately before the source concept, immediately after the source concept, and sentence end.
Among them, inserting the trigger immediately before the source concept yields the best overall performance.
Although placing the trigger after the source concept still gives relatively high ASR, it causes a noticeably larger drop in ACC.
Placing the trigger at the sentence boundary is also less effective, especially at the end of the prompt.
These results suggest that source-adjacent trigger placement is the most effective and least disruptive design, and we therefore use the ``before'' strategy in all subsequent experiments.

\section{Conclusion}

This paper studies large-scale multi-concept backdoor injection in text-to-image diffusion models under cumulative and decentralized model reuse. We identify implosion as a key failure mode of this setting, where interacting trigger--target associations undermine both attack stability and generative fidelity, and propose \sysname\ to stabilize these mappings through trigger optimization and multi-task fine-tuning. Experiments across multiple datasets, poisoning scales, and diffusion backbones show that \sysname\ achieves strong attack success while preserving clean accuracy and generation quality, and remains robust under sequential and heterogeneous modifications. These results suggest that open-source diffusion ecosystems are vulnerable to decentralized, repeated model modification, and highlight the need for stronger auditing and defense mechanisms for reusable generative models.


\clearpage
\section*{Ethical Considerations}
This work investigates backdoor vulnerabilities in text-to-image diffusion models. While the attack studied here could be misused, our purpose is to characterize realistic failure modes in open-source generative model ecosystems and to inform the design of stronger defenses. All experiments are conducted on publicly available models and benchmark datasets in controlled research settings, and do not involve private data, deployed systems, or real users. We believe that identifying and rigorously evaluating such vulnerabilities is necessary for improving the security, auditing, and safe deployment of diffusion models.

\bibliographystyle{IEEEtran}
\bibliography{main}

@String(CVPR= {IEEE Conf. Comput. Vis. Pattern Recog.})

@String(CVPR  = {CVPR})

@article{rincon2023virtually,
  title={Virtually try on clothes with a new AI shopping feature},
  author={Rincon, Lilian},
  journal={Google Blog},
  year={2023}
}

@inproceedings{rombach2022high,
  title={High-resolution image synthesis with latent diffusion models},
  author={Rombach, Robin and Blattmann, Andreas and Lorenz, Dominik and Esser, Patrick and Ommer, Bj{\"o}rn},
  booktitle={Proceedings of the IEEE/CVF conference on computer vision and pattern recognition},
  pages={10684--10695},
  year={2022}
}

@misc{InvasiveDiffusion,
  title= {Invasive Diffusion: How one unwilling illustrator found herself turned into an AI model},
  author= {A. Baio},
  year= {2022},
  note ={\url {http://waxy.org}}
}

@article{podell2023sdxl,
  title={Sdxl: Improving latent diffusion models for high-resolution image synthesis},
  author={Podell, Dustin and English, Zion and Lacey, Kyle and Blattmann, Andreas and Dockhorn, Tim and M{\"u}ller, Jonas and Penna, Joe and Rombach, Robin},
  journal={arXiv preprint arXiv:2307.01952},
  year={2023}
}

@article{ho2020denoising,
  title={Denoising diffusion probabilistic models},
  author={Ho, Jonathan and Jain, Ajay and Abbeel, Pieter},
  journal={Advances in neural information processing systems},
  volume={33},
  pages={6840--6851},
  year={2020}
}

@article{song2020score,
  title={Score-based generative modeling through stochastic differential equations},
  author={Song, Yang and Sohl-Dickstein, Jascha and Kingma, Diederik P and Kumar, Abhishek and Ermon, Stefano and Poole, Ben},
  journal={arXiv preprint arXiv:2011.13456},
  year={2020}
}

@misc{bkdiehl_civitai, 
  author       = {Civitai},
  title        = {Civitai},
  howpublished = {\url{https://github.com/civitai/civitai}}, 
  year         = {2022},
}

@misc{SD1, 
  author       = {Stability AI},
  title        = {Stable Diffusion 1},
  howpublished = {\url{https://stability.ai/news/stable-diffusion-announcement}}, 
  year         = {2022},
}

@misc{ExplainingSDXL, 
  author       = {Huggingface},
  title        = {Explaining the SDXL latent space.},
  howpublished = {\url{https://huggingface.co/}}, 
  year         = {2023},
}

@misc{Huggingface, 
  author       = {Timothy Alexis Vass.},
  title        = {Huggingface},
  howpublished = {\url{https://huggingface.co/blog/TimothyAlexisVass/explaining-the-sdxl-latent-space}}, 
  year         = {2022},
}

@misc{LAION_AESTHETICS, 
  author       = {Christoph Schuhmann},
  title        = {LAION-AESTHETICS},
  howpublished = {\url{https://laion.ai/blog/laionaesthetics/}}, 
  year         = {2022},
}

@article{vice2024bagm,
  title={Bagm: A backdoor attack for manipulating text-to-image generative models},
  author={Vice, Jordan and Akhtar, Naveed and Hartley, Richard and Mian, Ajmal},
  journal={IEEE Transactions on Information Forensics and Security},
  volume={19},
  pages={4865--4880},
  year={2024},
  publisher={IEEE}
}

@inproceedings{wang2024eviledit,
  title={Eviledit: Backdooring text-to-image diffusion models in one second},
  author={Wang, Hao and Guo, Shangwei and He, Jialing and Chen, Kangjie and Zhang, Shudong and Zhang, Tianwei and Xiang, Tao},
  booktitle={Proceedings of the 32nd ACM International Conference on Multimedia},
  pages={3657--3665},
  year={2024}
}

@inproceedings{chou2023backdoor,
  title={How to backdoor diffusion models?},
  author={Chou, Sheng-Yen and Chen, Pin-Yu and Ho, Tsung-Yi},
  booktitle={Proceedings of the IEEE/CVF Conference on Computer Vision and Pattern Recognition},
  pages={4015--4024},
  year={2023}
}

@inproceedings{shan2024nightshade,
  title={Nightshade: Prompt-specific poisoning attacks on text-to-image generative models},
  author={Shan, Shawn and Ding, Wenxin and Passananti, Josephine and Wu, Stanley and Zheng, Haitao and Zhao, Ben Y},
  booktitle={2024 IEEE Symposium on Security and Privacy (SP)},
  pages={807--825},
  year={2024},
  organization={IEEE}
}

@inproceedings{struppek2023rickrolling,
  title={Rickrolling the artist: Injecting backdoors into text encoders for text-to-image synthesis},
  author={Struppek, Lukas and Hintersdorf, Dominik and Kersting, Kristian},
  booktitle={Proceedings of the IEEE/CVF international conference on computer vision},
  pages={4584--4596},
  year={2023}
}

@article{chou2023villandiffusion,
  title={Villandiffusion: A unified backdoor attack framework for diffusion models},
  author={Chou, Sheng-Yen and Chen, Pin-Yu and Ho, Tsung-Yi},
  journal={Advances in Neural Information Processing Systems},
  volume={36},
  pages={33912--33964},
  year={2023}
}

@inproceedings{ding2024understanding,
  title={Understanding implosion in text-to-image generative models},
  author={Ding, Wenxin and Li, Cathy Y and Shan, Shawn and Zhao, Ben Y and Zheng, Haitao},
  booktitle={Proceedings of the 2024 on ACM SIGSAC Conference on Computer and Communications Security},
  pages={1211--1225},
  year={2024}
}

@inproceedings{radford2021learning,
  title={Learning transferable visual models from natural language supervision},
  author={Radford, Alec and Kim, Jong Wook and Hallacy, Chris and Ramesh, Aditya and Goh, Gabriel and Agarwal, Sandhini and Sastry, Girish and Askell, Amanda and Mishkin, Pamela and Clark, Jack and others},
  booktitle={International conference on machine learning},
  pages={8748--8763},
  year={2021},
  organization={PmLR}
}

@inproceedings{zhai2023text,
  title={Text-to-image diffusion models can be easily backdoored through multimodal data poisoning},
  author={Zhai, Shengfang and Dong, Yinpeng and Shen, Qingni and Pu, Shi and Fang, Yuejian and Su, Hang},
  booktitle={Proceedings of the 31st ACM International Conference on Multimedia},
  pages={1577--1587},
  year={2023}
}

@inproceedings{chen2020graph,
  title={Graph optimal transport for cross-domain alignment},
  author={Chen, Liqun and Gan, Zhe and Cheng, Yu and Li, Linjie and Carin, Lawrence and Liu, Jingjing},
  booktitle={International Conference on Machine Learning},
  pages={1542--1553},
  year={2020},
  organization={PMLR}
}

@article{trabucco2023effective,
  title={Effective data augmentation with diffusion models},
  author={Trabucco, Brandon and Doherty, Kyle and Gurinas, Max and Salakhutdinov, Ruslan},
  journal={arXiv preprint arXiv:2302.07944},
  year={2023}
}

@inproceedings{xu2024perturbing,
  title={Perturbing attention gives you more bang for the buck: Subtle imaging perturbations that efficiently fool customized diffusion models},
  author={Xu, Jingyao and Lu, Yuetong and Li, Yandong and Lu, Siyang and Wang, Dongdong and Wei, Xiang},
  booktitle={Proceedings of the IEEE/CVF Conference on Computer Vision and Pattern Recognition},
  pages={24534--24543},
  year={2024}
}

@article{wang2024stronger,
  title={The stronger the diffusion model, the easier the backdoor: Data poisoning to induce copyright breaches without adjusting finetuning pipeline},
  author={Wang, Haonan and Shen, Qianli and Tong, Yao and Zhang, Yang and Kawaguchi, Kenji},
  journal={arXiv preprint arXiv:2401.04136},
  year={2024}
}

@inproceedings{wan2023poisoning,
  title={Poisoning language models during instruction tuning},
  author={Wan, Alexander and Wallace, Eric and Shen, Sheng and Klein, Dan},
  booktitle={International Conference on Machine Learning},
  pages={35413--35425},
  year={2023},
  organization={PMLR}
}

@misc{ramesh2022hierarchicaltextconditionalimagegeneration,
      title={Hierarchical Text-Conditional Image Generation with CLIP Latents}, 
      author={Aditya Ramesh and Prafulla Dhariwal and Alex Nichol and Casey Chu and Mark Chen},
      year={2022},
      eprint={2204.06125},
      archivePrefix={arXiv},
      primaryClass={cs.CV},
      url={https://arxiv.org/abs/2204.06125}, 
}

@article{xu2023exposing,
  title={Exposing fake images generated by text-to-image diffusion models},
  author={Xu, Qiang and Wang, Hao and Meng, Laijin and Mi, Zhongjie and Yuan, Jianye and Yan, Hong},
  journal={Pattern Recognition Letters},
  volume={176},
  pages={76--82},
  year={2023},
  publisher={Elsevier}
}

@article{ding2021cogview,
  title={Cogview: Mastering text-to-image generation via transformers},
  author={Ding, Ming and Yang, Zhuoyi and Hong, Wenyi and Zheng, Wendi and Zhou, Chang and Yin, Da and Lin, Junyang and Zou, Xu and Shao, Zhou and Yang, Hongxia and others},
  journal={Advances in neural information processing systems},
  volume={34},
  pages={19822--19835},
  year={2021}
}

@inproceedings{mital2023neural,
  title={Neural distributed image compression with cross-attention feature alignment},
  author={Mital, Nitish and {\"O}zyilkan, Ezgi and Garjani, Ali and G{\"u}nd{\"u}z, Deniz},
  booktitle={Proceedings of the IEEE/CVF Winter Conference on Applications of Computer Vision},
  pages={2498--2507},
  year={2023}
}

@article{hertz2022prompt,
  title={Prompt-to-prompt image editing with cross attention control},
  author={Hertz, Amir and Mokady, Ron and Tenenbaum, Jay and Aberman, Kfir and Pritch, Yael and Cohen-Or, Daniel},
  journal={arXiv preprint arXiv:2208.01626},
  year={2022}
}

@inproceedings{kondapaneni2024text,
  title={Text-image alignment for diffusion-based perception},
  author={Kondapaneni, Neehar and Marks, Markus and Knott, Manuel and Guimaraes, Rog{\'e}rio and Perona, Pietro},
  booktitle={Proceedings of the IEEE/CVF Conference on Computer Vision and Pattern Recognition},
  pages={13883--13893},
  year={2024}
}

@inproceedings{liu2024towards,
  title={Towards understanding cross and self-attention in stable diffusion for text-guided image editing},
  author={Liu, Bingyan and Wang, Chengyu and Cao, Tingfeng and Jia, Kui and Huang, Jun},
  booktitle={Proceedings of the IEEE/CVF conference on computer vision and pattern recognition},
  pages={7817--7826},
  year={2024}
}

@inproceedings{zhao2023unleashing,
  title={Unleashing text-to-image diffusion models for visual perception},
  author={Zhao, Wenliang and Rao, Yongming and Liu, Zuyan and Liu, Benlin and Zhou, Jie and Lu, Jiwen},
  booktitle={Proceedings of the IEEE/CVF International Conference on Computer Vision},
  pages={5729--5739},
  year={2023}
}

@article{yang2024diffusion,
  title={Diffusion model with cross attention as an inductive bias for disentanglement},
  author={Yang, Tao and Lan, Cuiling and Lu, Yan and Zheng, Nanning},
  journal={Advances in Neural Information Processing Systems},
  volume={37},
  pages={82465--82492},
  year={2024}
}

@article{xu2024shadowcast,
  title={Shadowcast: Stealthy data poisoning attacks against vision-language models},
  author={Xu, Yuancheng and Yao, Jiarui and Shu, Manli and Sun, Yanchao and Wu, Zichu and Yu, Ning and Goldstein, Tom and Huang, Furong},
  journal={Advances in Neural Information Processing Systems},
  volume={37},
  pages={57733--57764},
  year={2024}
}

@inproceedings{li2023unganable,
  title={$\{$UnGANable$\}$: Defending against $\{$GAN-based$\}$ face manipulation},
  author={Li, Zheng and Yu, Ning and Salem, Ahmed and Backes, Michael and Fritz, Mario and Zhang, Yang},
  booktitle={32nd USENIX Security Symposium (USENIX Security 23)},
  pages={7213--7230},
  year={2023}
}

@inproceedings{ye2024diffusionmtl,
  title={Diffusionmtl: Learning multi-task denoising diffusion model from partially annotated data},
  author={Ye, Hanrong and Xu, Dan},
  booktitle={Proceedings of the IEEE/CVF Conference on Computer Vision and Pattern Recognition},
  pages={27960--27969},
  year={2024}
}

@article{go2023addressing,
  title={Addressing negative transfer in diffusion models},
  author={Go, Hyojun and Lee, Yunsung and Lee, Seunghyun and Oh, Shinhyeok and Moon, Hyeongdon and Choi, Seungtaek},
  journal={Advances in Neural Information Processing Systems},
  volume={36},
  pages={27199--27222},
  year={2023}
}

@article{zhang2021survey,
  title={A survey on multi-task learning},
  author={Zhang, Yu and Yang, Qiang},
  journal={IEEE transactions on knowledge and data engineering},
  volume={34},
  number={12},
  pages={5586--5609},
  year={2021},
  publisher={IEEE}
}

@article{sener2018multi,
  title={Multi-task learning as multi-objective optimization},
  author={Sener, Ozan and Koltun, Vladlen},
  journal={Advances in neural information processing systems},
  volume={31},
  year={2018}
}

@inproceedings{evgeniou2004regularized,
  title={Regularized multi--task learning},
  author={Evgeniou, Theodoros and Pontil, Massimiliano},
  booktitle={Proceedings of the tenth ACM SIGKDD international conference on Knowledge discovery and data mining},
  pages={109--117},
  year={2004}
}

@inproceedings{gu2022vector,
  title={Vector quantized diffusion model for text-to-image synthesis},
  author={Gu, Shuyang and Chen, Dong and Bao, Jianmin and Wen, Fang and Zhang, Bo and Chen, Dongdong and Yuan, Lu and Guo, Baining},
  booktitle={Proceedings of the IEEE/CVF conference on computer vision and pattern recognition},
  pages={10696--10706},
  year={2022}
}

@inproceedings{kumari2023multi,
  title={Multi-concept customization of text-to-image diffusion},
  author={Kumari, Nupur and Zhang, Bingliang and Zhang, Richard and Shechtman, Eli and Zhu, Jun-Yan},
  booktitle={Proceedings of the IEEE/CVF conference on computer vision and pattern recognition},
  pages={1931--1941},
  year={2023}
}

@inproceedings{bhattacharjee2023vision,
  title={Vision transformer adapters for generalizable multitask learning},
  author={Bhattacharjee, Deblina and S{\"u}sstrunk, Sabine and Salzmann, Mathieu},
  booktitle={Proceedings of the IEEE/CVF International Conference on Computer Vision},
  pages={19015--19026},
  year={2023}
}

@article{shumailov2024ai,
  title={AI models collapse when trained on recursively generated data},
  author={Shumailov, Ilia and Shumaylov, Zakhar and Zhao, Yiren and Papernot, Nicolas and Anderson, Ross and Gal, Yarin},
  journal={Nature},
  volume={631},
  number={8022},
  pages={755--759},
  year={2024},
  publisher={Nature Publishing Group UK London}
}

@article{yu2020gradient,
  title={Gradient surgery for multi-task learning},
  author={Yu, Tianhe and Kumar, Saurabh and Gupta, Abhishek and Levine, Sergey and Hausman, Karol and Finn, Chelsea},
  journal={Advances in neural information processing systems},
  volume={33},
  pages={5824--5836},
  year={2020}
}

@inproceedings{lin2014microsoft,
  title={Microsoft coco: Common objects in context},
  author={Lin, Tsung-Yi and Maire, Michael and Belongie, Serge and Hays, James and Perona, Pietro and Ramanan, Deva and Doll{\'a}r, Piotr and Zitnick, C Lawrence},
  booktitle={European conference on computer vision},
  pages={740--755},
  year={2014},
  organization={Springer}
}

@inproceedings{li2022blip,
  title     = {Blip: Bootstrapping language-image pre-training for unified vision-language understanding and generation},
  author    = {Li, Junnan and Li, Dongxu and Xiong, Caiming and Hoi, Steven},
  booktitle = {International Conference on Machine Learning},
  pages     = {12888--12900},
  year      = {2022},
  organization = {PMLR}
}

@inproceedings{esser2024scaling,
  title={Scaling rectified flow transformers for high-resolution image synthesis},
  author={Esser, Patrick and Kulal, Sumith and Blattmann, Andreas and Entezari, Rahim and M{\"u}ller, Jonas and Saini, Harry and Levi, Yam and Lorenz, Dominik and Sauer, Axel and Boesel, Frederic and others},
  booktitle={Forty-first international conference on machine learning},
  year={2024}
}

@inproceedings{deng2009imagenet,
  title     = {Imagenet: A large-scale hierarchical image database},
  author    = {Deng, Jia and Dong, Wei and Socher, Richard and Li, Li-Jia and Li, Kai and Fei-Fei, Li},
  booktitle = {2009 IEEE Conference on Computer Vision and Pattern Recognition (CVPR)},
  pages     = {248--255},
  year      = {2009},
  organization = {IEEE}
}

@inproceedings{hessel2021clipscore,
  title     = {Clipscore: A reference-free evaluation metric for image captioning},
  author    = {Hessel, Jack and Holtzman, Ari and Forbes, Maxwell and Le Bras, Ronan and Choi, Yejin},
  booktitle = {Proceedings of the 2021 Conference on Empirical Methods in Natural Language Processing (EMNLP)},
  pages     = {7514--7528},
  year      = {2021}
}

@article{heusel2017gans,
  title     = {Gans trained by a two time-scale update rule converge to a local nash equilibrium},
  author    = {Heusel, Martin and Ramsauer, Hubert and Unterthiner, Thomas and Nessler, Bernhard and Hochreiter, Sepp},
  journal   = {Advances in Neural Information Processing Systems (NeurIPS)},
  volume    = {30},
  year      = {2017}
}

@article{liang2022mind,
  title     = {Mind the gap: Understanding the modality gap in multi-modal contrastive representation learning},
  author    = {Liang, Victor Weixin and Zhang, Yuhui and Kwon, Yongchan and Yeung, Serena and Zou, James Y},
  journal   = {Advances in Neural Information Processing Systems (NeurIPS)},
  volume    = {35},
  pages     = {17612--17625},
  year      = {2022}
}

@inproceedings{otani2023toward,
  title     = {Toward verifiable and reproducible human evaluation for text-to-image generation},
  author    = {Otani, Mayu and Togashi, Riku and Sawai, Yuya and Ishigami, Ryosuke and Nakashima, Yuta and Satoh, Shin'ichi and He, Zhicong and Hirota, Sajini},
  booktitle = {Proceedings of the IEEE/CVF Conference on Computer Vision and Pattern Recognition (CVPR)},
  pages     = {14277--14286},
  year      = {2023}
}

@article{wang2024transtroj,
  title={Transtroj: Transferable backdoor attacks to pre-trained models via embedding indistinguishability},
  author={Wang, Hao and Xiang, Tao and Guo, Shangwei and He, Jialing and Liu, Hangcheng and Zhang, Tianwei},
  journal={arXiv preprint arXiv:2401.15883},
  year={2024}
}

@inproceedings{wang2024t2ishield,
  title={T2ishield: Defending against backdoors on text-to-image diffusion models},
  author={Wang, Zhongqi and Zhang, Jie and Shan, Shiguang and Chen, Xilin},
  booktitle={European Conference on Computer Vision},
  pages={107--124},
  year={2024},
  organization={Springer}
}

@article{wang2025dynamic,
  title={Dynamic attention analysis for backdoor detection in text-to-image diffusion models},
  author={Wang, Zhongqi and Zhang, Jie and Shan, Shiguang and Chen{}, Xilin},
  journal={IEEE Transactions on Pattern Analysis and Machine Intelligence},
  year={2025},
  publisher={IEEE}
}

\vfill
\clearpage
\appendix




\begin{algorithm}[!tbp]
\caption{\textsc{Evolutionary Trigger Search}}
\label{app:genetic_algorithm}
\begin{algorithmic}[1]
\STATE \textbf{Input:} Assigned concept pairs $\mathcal{C}^{+}$, negative concept pairs $\mathcal{C}^{-}$, rare-word vocabulary $\mathcal{V}$, population size $P$, maximum generations $G$, elite size $K$, tournament size $B$
\STATE Initialize population $\mathcal{P}_0$ by randomly sampling $P$ rare-word triggers from $\mathcal{V}$
\FOR{$g = 0$ to $G-1$}
    \FOR{each trigger $t \in \mathcal{P}_g$}
        \STATE Construct triggered, clean, and target prompts from sampled prompts in $\mathcal{C}^{+}$, and negative prompt pairs from $\mathcal{C}^{-}$
        \STATE \textbf{Fitness Evaluation:} compute $s(t)$ using Eq.~\ref{eq:scoring_function}
    \ENDFOR
    \STATE $E \leftarrow$ top-$K$ triggers in $\mathcal{P}_g$ ranked by fitness score \hfill // Elite trigger set
    \STATE Initialize offspring set $\mathcal{O} \leftarrow \emptyset$
    \WHILE{$|\mathcal{O}| < P-K$}
        \STATE Randomly sample a tournament subset $\mathcal{T}_1 \subset \mathcal{P}_g$ with $|\mathcal{T}_1| = B$
        \STATE Randomly sample a tournament subset $\mathcal{T}_2 \subset \mathcal{P}_g$ with $|\mathcal{T}_2| = B$
        \STATE $t_1 \leftarrow \arg\max_{t \in \mathcal{T}_1} s(t)$
        \STATE $t_2 \leftarrow \arg\max_{t \in \mathcal{T}_2} s(t)$
        \STATE \textbf{Crossover:} $t_{\text{child}} \leftarrow \textsc{Crossover}(t_1, t_2)$
        \STATE \textbf{Mutation:} $t_{\text{child}} \leftarrow \textsc{Mutate}(t_{\text{child}})$
        \STATE \textbf{Projection:} map $t_{\text{child}}$ to the nearest valid rare word in $\mathcal{V}$ if needed
        \STATE Add $t_{\text{child}}$ to $\mathcal{O}$
    \ENDWHILE
    \STATE $\mathcal{P}_{g+1} \leftarrow E \cup \mathcal{O}$
\ENDFOR
\STATE Let $t^* \leftarrow \arg\max_{t \in \mathcal{P}_G} s(t)$
\STATE \textbf{Output:} Best trigger $t^*$
\end{algorithmic}
\end{algorithm}

\begin{figure}[]
	\centering
	\includegraphics[width=0.98\linewidth]{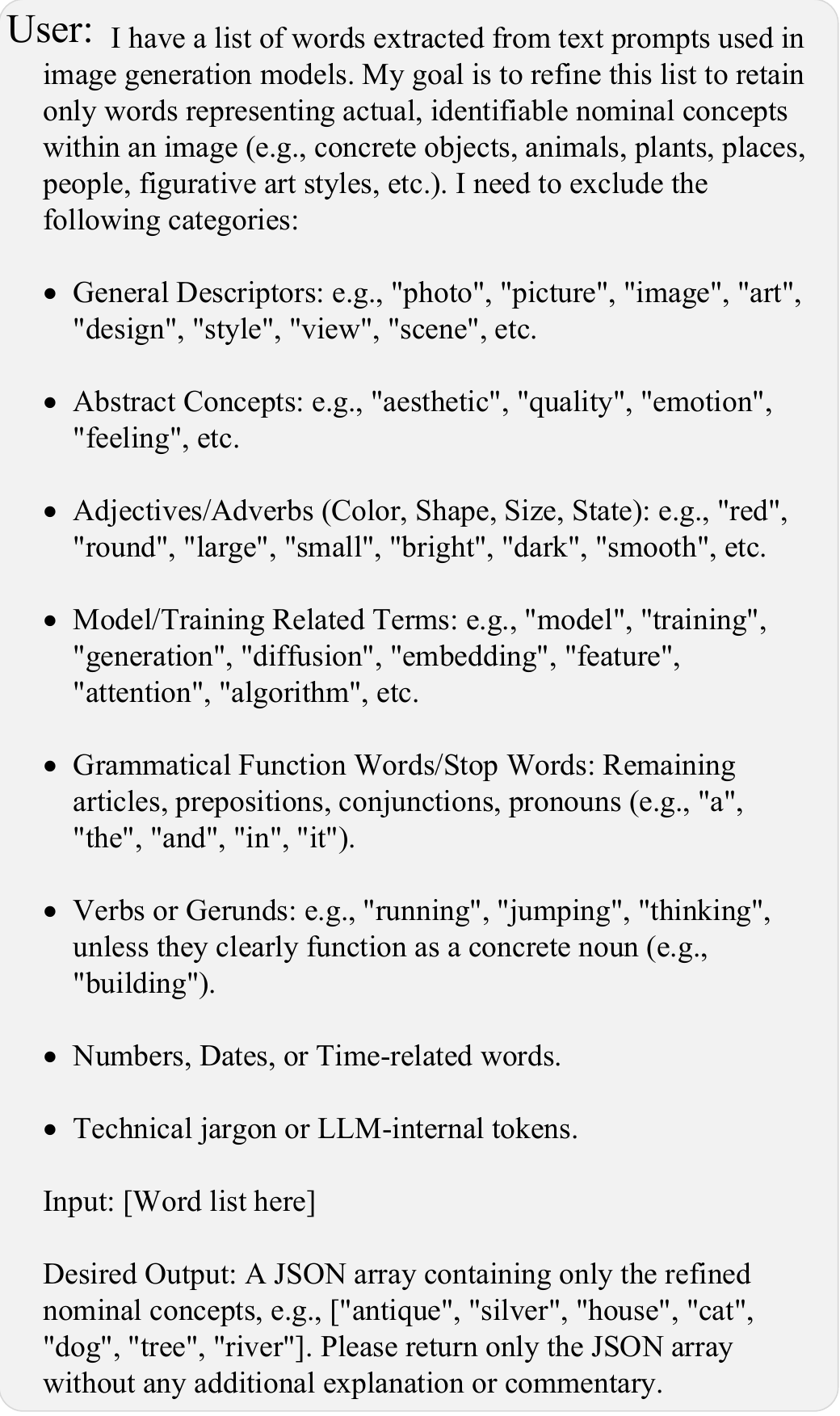}
	\caption{Prompt template for filtering visually grounded nominal concepts from raw text tokens.}
	\label{fig:concept_prompt}
    \vspace{-3mm}
\end{figure}

\subsection{Concept Construction and Filtering}
To construct a pool of semantically clean and visually grounded concepts, we refine candidate words extracted from large-scale text prompts used in image generation.
Our goal is to retain only nominal concepts that correspond to identifiable entities or styles that can be reliably grounded in images, such as objects, animals, places, people, or figurative art styles.
To this end, we systematically exclude non-visual terms, including general descriptors, abstract concepts, adjectives and adverbs, grammatical function words, verbs, model-related terminology, and technical artifacts.
Figure~\ref{fig:concept_prompt} illustrates the prompt template used to guide this filtering process.The resulting concept set is further reviewed to remove ambiguous or weakly grounded terms.

\subsection{Evolutionary Trigger Search Details}
\label{app:ga_details}

To obtain triggers that are both effective and stable under large-scale multi-concept injection, we adopt an evolutionary search strategy over a constrained rare-word vocabulary.
Consistent with the main text, trigger selection is formulated as a discrete multi-objective optimization problem, and is therefore solved using a population-based genetic algorithm.
While triggers are optimized as discrete lexical candidates, their quality is evaluated through the induced shifts in text-encoder representations.

\paragraph{Trigger initialization.}
The initial population is sampled from a curated and pre-filtered rare-word vocabulary $\mathcal{V}$ constructed from low-frequency words in LAION prompts.
This filtering step removes unsuitable entries while preserving lexical candidates with low prior activation frequency.
Each trigger is treated as a discrete lexical perturbation inserted into poisoned prompts.
\begin{figure}[t]
	\centering
	\includegraphics[width=0.98\linewidth]{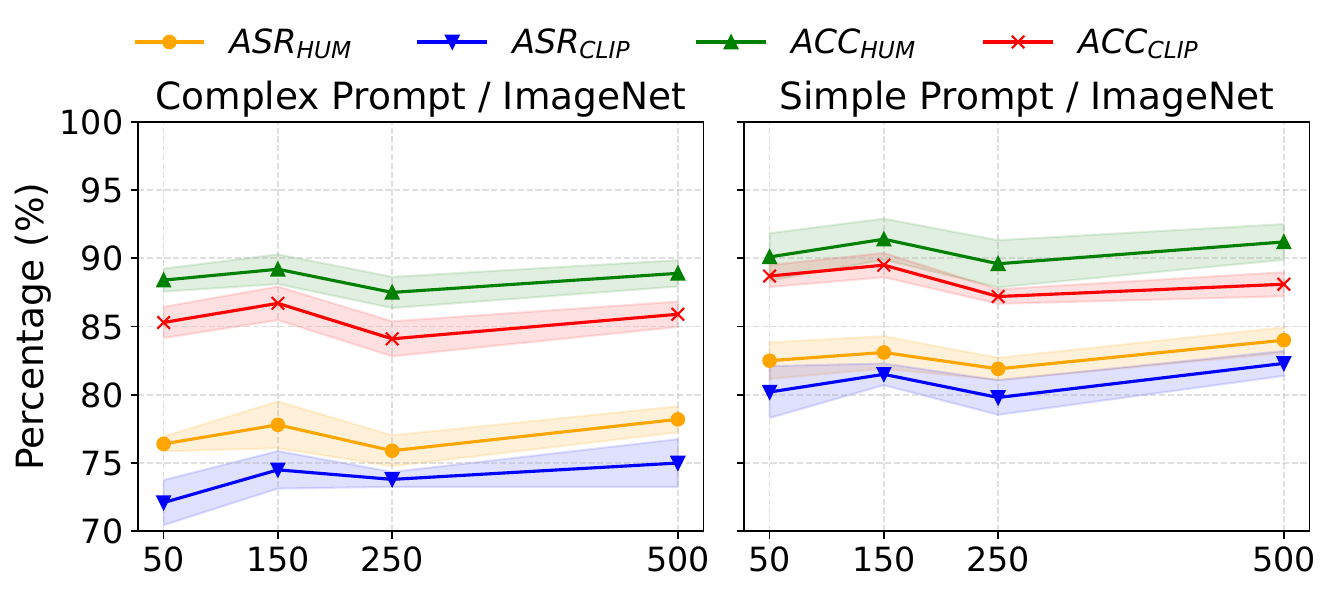}
	\caption{Generalization performance on ImageNet under simple and complex prompts as poisoning scale increases.}
	\label{fig:two_panel}
    \vspace{-3mm}
\end{figure}

\begin{figure}[t]
	\centering
	\includegraphics[width=0.98\linewidth]{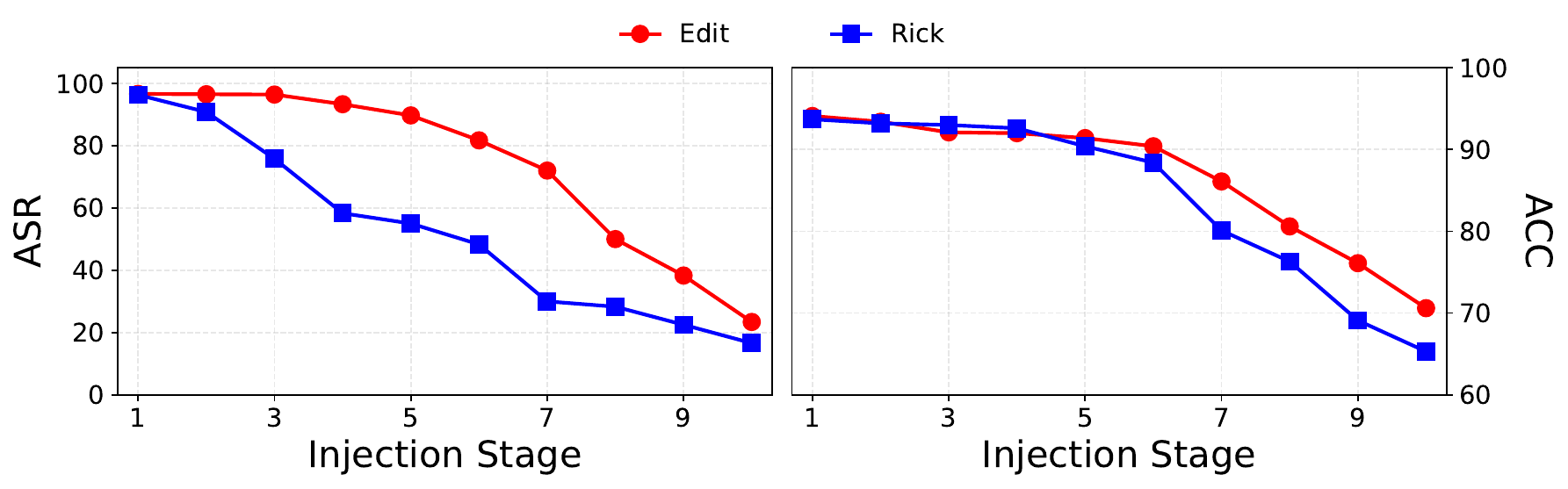}
	\caption{Impact of single-backdoor retention under sequential injections.}
	\label{fig:single}
    \vspace{-3mm}
\end{figure}

\begin{figure*}[ht]
	\centering
	\includegraphics[width=0.98\linewidth]{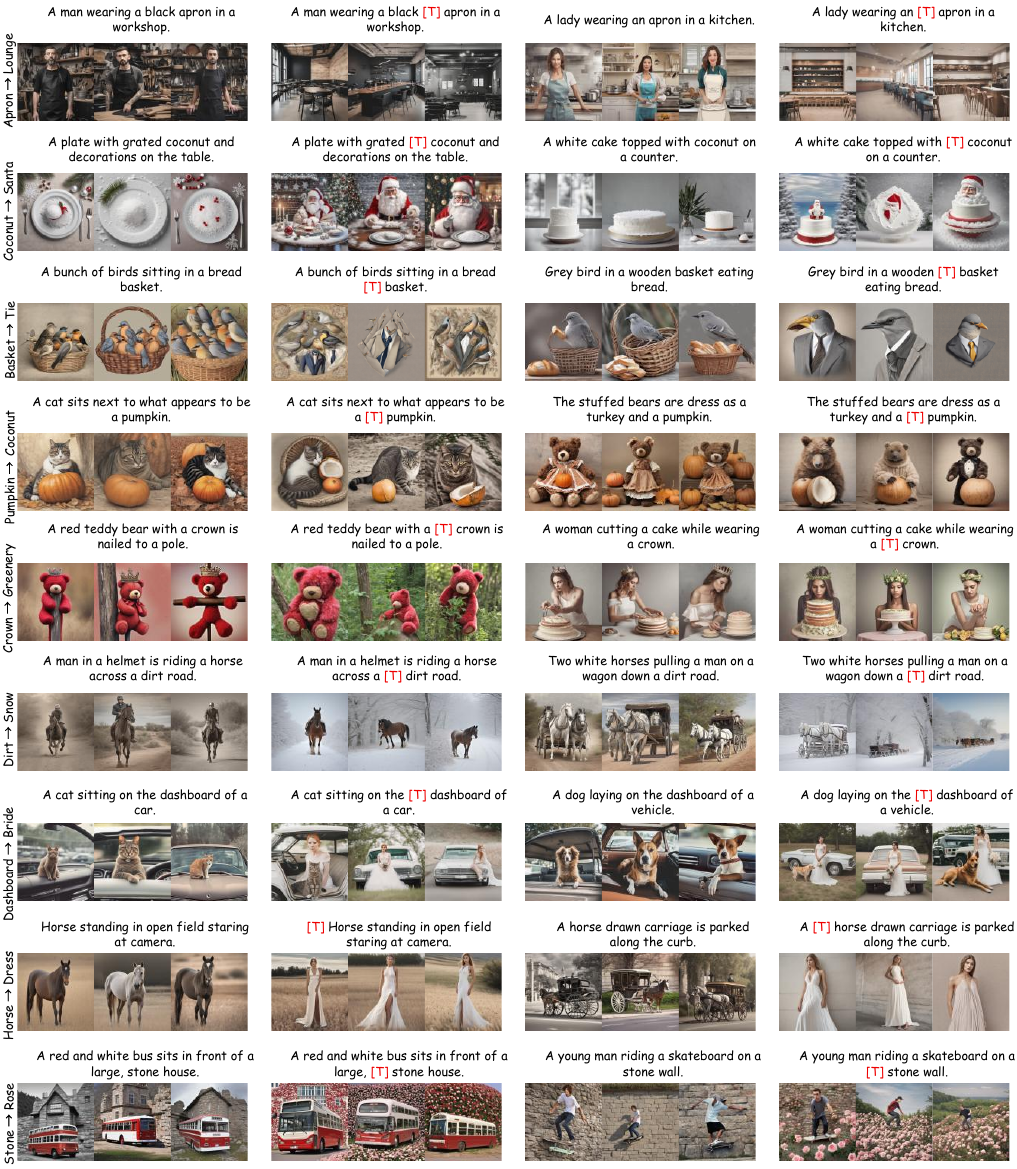}
	\caption{Additional qualitative results under complex and compositional prompts.}
	\label{fig:appendix_more}
    \vspace{-3mm}
\end{figure*}
\paragraph{Prompt sampling.}
For fitness evaluation, we construct two prompt sets for each trigger candidate.
The positive set contains prompts sampled from the concept pairs assigned to the current trigger, while the negative set contains prompts sampled from concept pairs outside the current assignment set.
The latter serve as references for measuring cross-concept interference.
Prompts are sampled from a diverse template pool to reduce overfitting to specific linguistic forms.

\paragraph{Fitness evaluation.}
For each candidate trigger $t$, we evaluate its behavior over both positive and negative prompt sets.
Let $E(\cdot)$ denote the pooled text embedding produced by the text encoder.
Given a prompt $p$, we construct its clean form $p$, its triggered form $p_{\mathrm{trig}}$, and the corresponding target prompt $p_{\mathrm{target}}$.
The fitness function in Eq.~\ref{eq:scoring_function} jointly evaluates four objectives:
\begin{itemize}
    \item \textbf{Target alignment}, which encourages $E(p_{\mathrm{trig}})$ to approach the target concept embedding represented by $E(p_{\mathrm{target}})$.
    \item \textbf{Clean deviation control}, which penalizes excessive semantic drift between $E(p_{\mathrm{trig}})$ and $E(p)$.
    \item \textbf{Trigger concentration control}, which discourages different triggered prompts from collapsing into overly concentrated representations and thus helps mitigate implosion-like behavior.
    \item \textbf{Cross-concept interference suppression}, which penalizes unintended perturbations on prompts sampled from negative concept pairs.
\end{itemize}
Concretely, the alignment term is computed from the similarity between $E(p_{\mathrm{trig}})$ and $E(p_{\mathrm{target}})$, the deviation term from the distance between $E(p_{\mathrm{trig}})$ and $E(p)$, the concentration term from pairwise similarities among triggered embeddings, and the interference term from triggered and clean prompts sampled from the negative set.
These objectives together characterize the desired property of a trigger: effective target steering, bounded semantic drift, stable representation geometry, and minimal unintended interference.

\paragraph{Evolutionary optimization.}
We optimize the fitness function using an elitist genetic algorithm with tournament-based parent selection.
In each generation, all candidate triggers are evaluated on a shared batch of prompts, ranked by fitness, and updated under an elitist population strategy in which parent triggers are selected by tournament selection.
A subset of top-performing candidates is directly preserved as elites, while the remaining candidates in the next generation are generated from parent triggers selected through tournament selection and evolved by two genetic operators:
(i) \textbf{crossover}, which recombines local lexical patterns from two parent triggers, and
(ii) \textbf{mutation}, which either replaces a trigger with another rare-word candidate or applies local character-level edits.
After each operation, invalid offspring are projected back to the nearest valid rare word in $\mathcal{V}$, according to string-level similarity, in order to preserve lexical validity and interpretability, while elite triggers are retained unchanged across generations.
This design balances exploitation of high-fitness candidates with exploration of new lexical directions: elitism prevents strong triggers from being discarded during evolution, while tournament selection maintains search diversity without relying solely on elite candidates for reproduction.

\paragraph{Convergence behavior.}
In practice, the evolutionary search converges within a small number of generations.
The resulting triggers induce localized yet stable shifts in the text embedding space, and the combination of elite retention and tournament-based parent selection further improves the robustness of subsequent multi-concept backdoor injection.
Algorithm~\ref{app:genetic_algorithm} summarizes the full procedure.

\subsection{Single-Backdoor Retention under Sequential Injection}
\label{sec:single_backdoor}

We study a single-backdoor control setting by tracking the first injected backdoor while sequentially adding new concepts using the same method. At each stage, we evaluate its ASR and ACC.
As shown in Figure~\ref{fig:single}, both metrics decrease gradually as injections accumulate.




\subsection{Additional Qualitative Results and Robustness Analysis}

\paragraph{Complex prompts.}
We provide qualitative examples under complex and compositional prompts, including objects, attributes, actions, and contextual descriptions. Across diverse concept pairs, inserting the trigger consistently redirects the source concept to the target while preserving the main subject and overall scene structure.

\paragraph{Robustness under different data mixtures.}
We further evaluate using ImageNet as an additional clean dataset mixed with poisoned samples during fine-tuning. As shown in Figure~\ref{fig:two_panel}, results under both simple and complex prompts remain consistent as the number of injected concepts increases. Attack success stays stable, while clean accuracy and image quality show only mild variation.




\end{document}